\documentclass[aps,pra,reprint,longbibliography,superscriptaddress]{revtex4-1}
\usepackage{mathtools}
\usepackage{bbold}
\usepackage{physics}
\usepackage{amsmath}
\usepackage{amsfonts}
\usepackage{xcolor}

\begin{document}

\title{Adiabatic pumping and transport in the Sierpinski-Hofstadter model}

\author{Saswat Sarangi}
\affiliation{Max-Planck-Institut f\"{u}r Physik komplexer Systeme,
D-01187 Dresden, Germany}

\author{Anne E. B. Nielsen}
\affiliation{Department of Physics and Astronomy,
Aarhus University, DK-8000 Aarhus C, Denmark}

\begin{abstract}
Topological phases have been reported on self-similar structures in the presence of a perpendicular magnetic field. Here, we present an understanding of these phases from a perspective of spectral flow and charge pumping. We study the Harper-Hofstadter model on self-similar structures constructed from the Sierpinski gasket. We numerically investigate the spectral flow and the associated charge pumping when a flux tube is inserted through the structure and the flux through the tube is varied adiabatically. We find that the nature of the spectral flow is qualitatively different from that of translationally invariant non-interacting systems with a perpendicular magnetic field. We show that the instantaneous eigenspectra can be used to understand the quantization of the charge pumped over a cycle, and hence to understand the topological character of the system. We show the correspondence between the local contributions to the Hall conductivity and the spectral flow of the edge-like states. We also show that the edge-like states can be approximated by eigenstates of the discrete angular-momentum operator, their chiral nature being a consequence of this.
\end{abstract}

\maketitle

\section{Introduction}
The study of topologically non-trivial phases is an important research area in condensed matter physics. In the context of non-interacting systems, these phases are well understood and classified in the presence of translational symmetry \cite{Schnyder2008,Chiu2016,Ludwig2015,Slager2013,Kane2005,Kane2005a,Bernevig2006,Fu2007,Fu2011}.
Also, in the absence of translational symmetry, topologically non-trivial phases have been reported in amorphous solids and quasicrystalline systems which retain the notion of a well-defined bulk and edge \cite{Adhip2017,Mitchell2018,Duncan2020}.
These  phases are identified by the presence of their signature robust edge states and are characterized by respective topological invariants.
\newline

In recent years, the study of topologically non-trivial phases in systems which lack the notion of a well-defined bulk and edge, has gained interest. Self-similar structures like finite truncations of the Sierpinski carpet and the Sierpinski gasket have been studied in the presence of a uniform perpendicular magnetic field \cite{Brzezinska2018,Fremling2019,Iliasov2020,Fischer2021}. Also, generalized two-orbital Bernevig-Hughes-Zhang models have been studied on such structures \cite{Adhip2018,Sarangi2021}. Several non-trivial phases have been reported in such systems. These phases seem to be identified with the presence of gapless `edge-like' states which are chiral in nature and are localized around each of the intrinsic `holes' present in these structures. Also, the Hall conductivity in such phases is shown to be quantized and robust to small disorders. 
\newline

Although, topological phase diagrams of some well-known models on self-similar structures are present in the literature, only a limited microscopic understanding of  such phases is available at present. For example, in the case of self-similar structures, understanding of these topologically non-trivial phases in terms of winding of the eigenstates over some manifold, analogous to the winding of the Bloch states in the $k$-space for translationally invariant non-interacting systems, is not present at the moment. Given the lack of an `eigenstate winding' perspective for self-similar structures, we use the perspective of adiabatic charge pumping in this work to understand the emergence of topology in self-similar systems and the quantization of real-space indices. Adiabatic pumping in translationally invariant non-interacting systems have been thoroughly studied. These systems form multiple magnetic bands (or Landau levels in the continuum case) when subjected to a perpendicular magnetic field. When additional flux is threaded through the system using a thin long solenoid, some states flow across the band-gap from one band to another. The Chern number can be expressed as the number of such states flowing across the band-gap. However, it is presently unclear if it is possible to directly translate these ideas and results over to the case of self-similar systems.
\newline

From our study, we find that the nature of the spectral flow is qualitatively different from that of translationally invariant non-interacting systems with perpendicular magnetic field. In this case, the spectral flow happens throughout the eigen-spectra as opposed to the case with translationally invariant systems where spectral flow is observed across the band gap. We find that the position of the flux-tube plays an important role in determining the states undergoing spectral flow. We show that the charge pumped is quantized in the adiabatic limit, irrespective of the position of the flux-tube.  We show that the instantaneous eigenspectra can be used to understand the quantization of the charge pumped over a cycle, making it a diagnostic tool to study the topological character of self-similar systems. We also explicitly calculate the local Hall conductivity of the system. We show the correspondence between the local contributions to the Hall conductivity and the spectral flow of the edge-like states. We also show that the edge-like states can be isolated from its degenerate group of states by tuning the flux through the flux-tube and find that they can be approximated by eigenstates of the discrete angular-momentum operator, their chiral nature being a consequence of this.
\newline

This article is organized in the following way. In Sec.\ \ref{setup}, we define the system by defining the model and the self-similar structures which are studied in this work. In Sec.\ \ref{Flux_pumping}, we describe the adiabatic charge pumping in the system by showing the instantaneous eigenspectra and by describing the quantization of the associated charge pumping. In Sec.\ \ref{prop_edge_states}, we study the properties of individual edge-like states. In Sec.\ \ref{sec_hall_cond}, we calculate the Hall conductivity, show its local contributions, and show its topological character by showing its quantization and robustness to disorder. Finally, we present a summary and discuss the outlook of this work in Sec.\ \ref{summary}.

\section{The setup}\label{setup}
\begin{figure}
\includegraphics[scale=0.16]{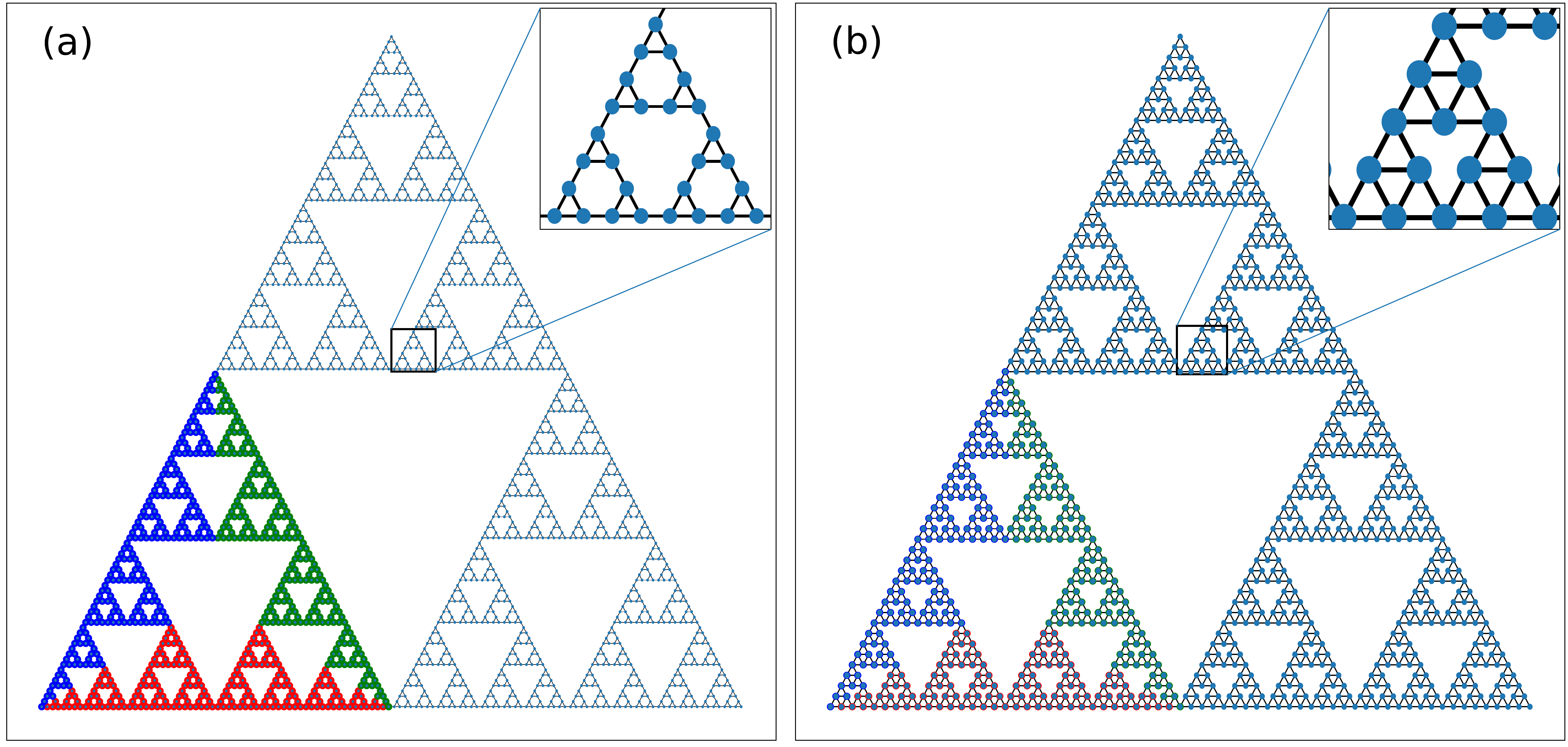}
\caption{(a) SG-3 with $g=7$, and (b) SG-4 with $g=7$. For the calculation of the real space Chern number using Eq.\ \eqref{KitaevChernNumber}, we choose a subsection of the system and divide it into three partitions. The partitions are shown with red, green and blue and the projectors onto these partitions are labeled $\mathbb{A}$, $\mathbb{B},$ and $\mathbb{C}$ respectively.}
\label{Syst_&_partition}
\end{figure}

We aim to study non-interacting spinless fermions on self-similar structures in the presence of a uniform magnetic field. The structures we consider are embedded in two dimensions and the magnetic field is perpendicular to the embedding surface. In this work, we consider two different discretizations of the Sierpinski gasket (SG), namely SG-3 and SG-4. These structures are constructed by discretizing the recursive generation scheme of the Sierpinski Gasket. Finite truncations of such structures are said to be of generation $g$ if the recursion scheme is truncated after the $g^{th}$ iteration. The detailed construction of these structures is mentioned in Ref.\ \cite{Sarangi2021}. Examples of finite generation of SG-3 and SG-4 are shown in Fig.\ \ref{Syst_&_partition}. We choose to primarily look at these two structures because of the relatively slow growth in the system size $N$ as a function of the generation $g$. For example, $N=3^{g}$ for SG-3 and $N=(3^{g}+3)/2$ for SG-4, whereas $N=8^{g}$ for a self-similar structure constructed from the Sierpinski Carpet. This makes it comparatively easier to numerically access higher generations and reach closer to the limiting fractional dimension for SG-3 and SG-4.
\newline

The Hamiltonian for the system is  the Harper-Hofstadter Hamiltonian given by:
\begin{equation}\label{HH}
H=-\sum_{<jk>}e^{-\text{i}\theta_{jk}}c^{\dagger}_{j}c_{k}+h.c.
\end{equation}
where $j,k$ are the labels for the sites positioned at $\vec{r}_j$ and $\vec{r}_k$, $<>$ denotes the nearest neighbors, and $\theta_{jk}=(1/\phi_0)\int_{\vec{r}_j}^{\vec{r}_j}\textbf{A} \cdot \textbf{dl}$ denotes the Peierls phase with the flux quantum $\phi_0=h/e$. Here, $\textbf{A}$ is the associated magnetic vector potential. For all the numerics, we have used the Landau gauge, $\textbf{A}=(0,Bx,0)$. We have parametrized the magnetic strength by $B=2\pi{\phi}/(\sqrt{3}a^2/4)$, where $a$ is the distance between nearest neighbor sites and $2\pi\phi$ is the flux piercing through the smallest triangles of the structures. 
\newline
 
\begin{figure}
\includegraphics[scale=0.163]{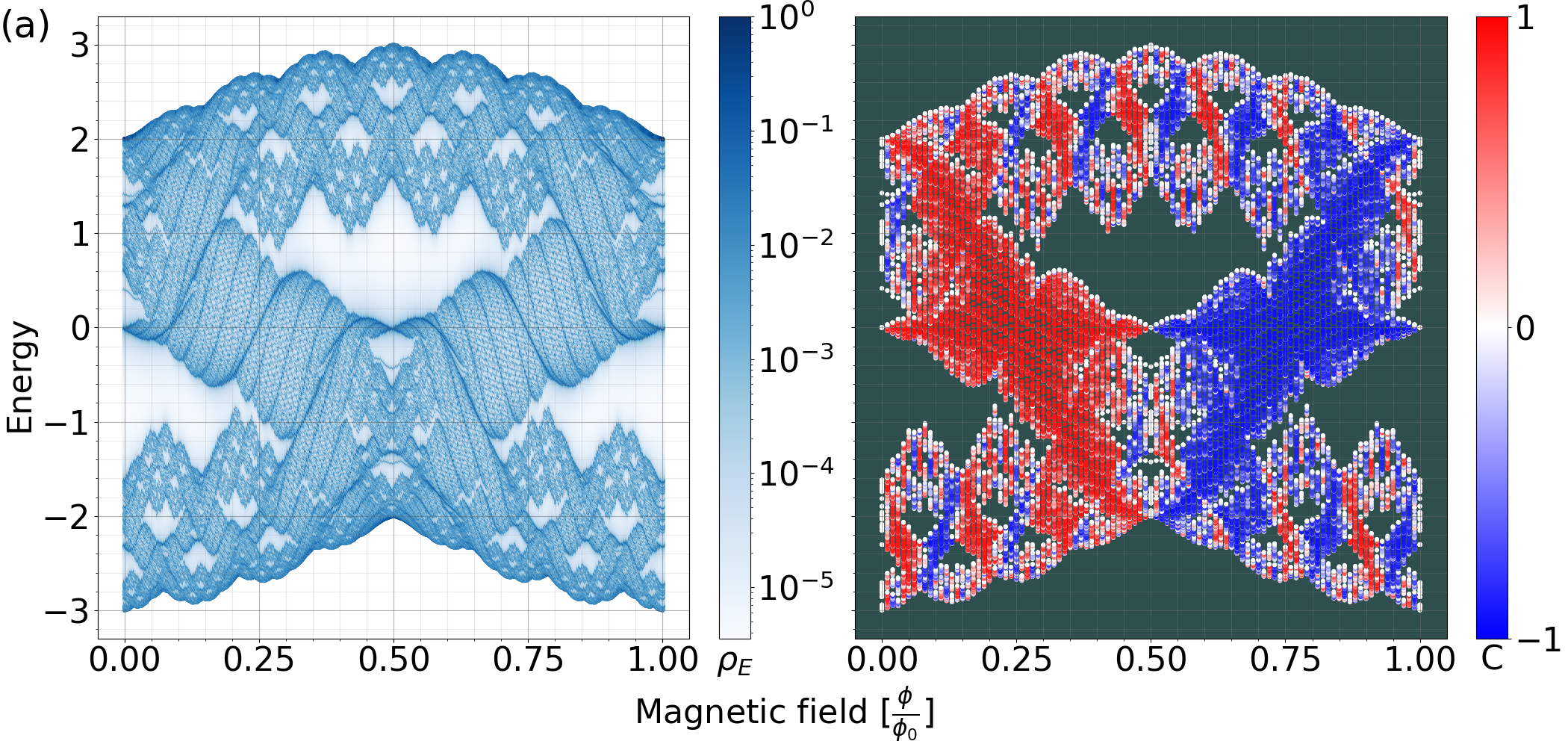}
\includegraphics[scale=0.163]{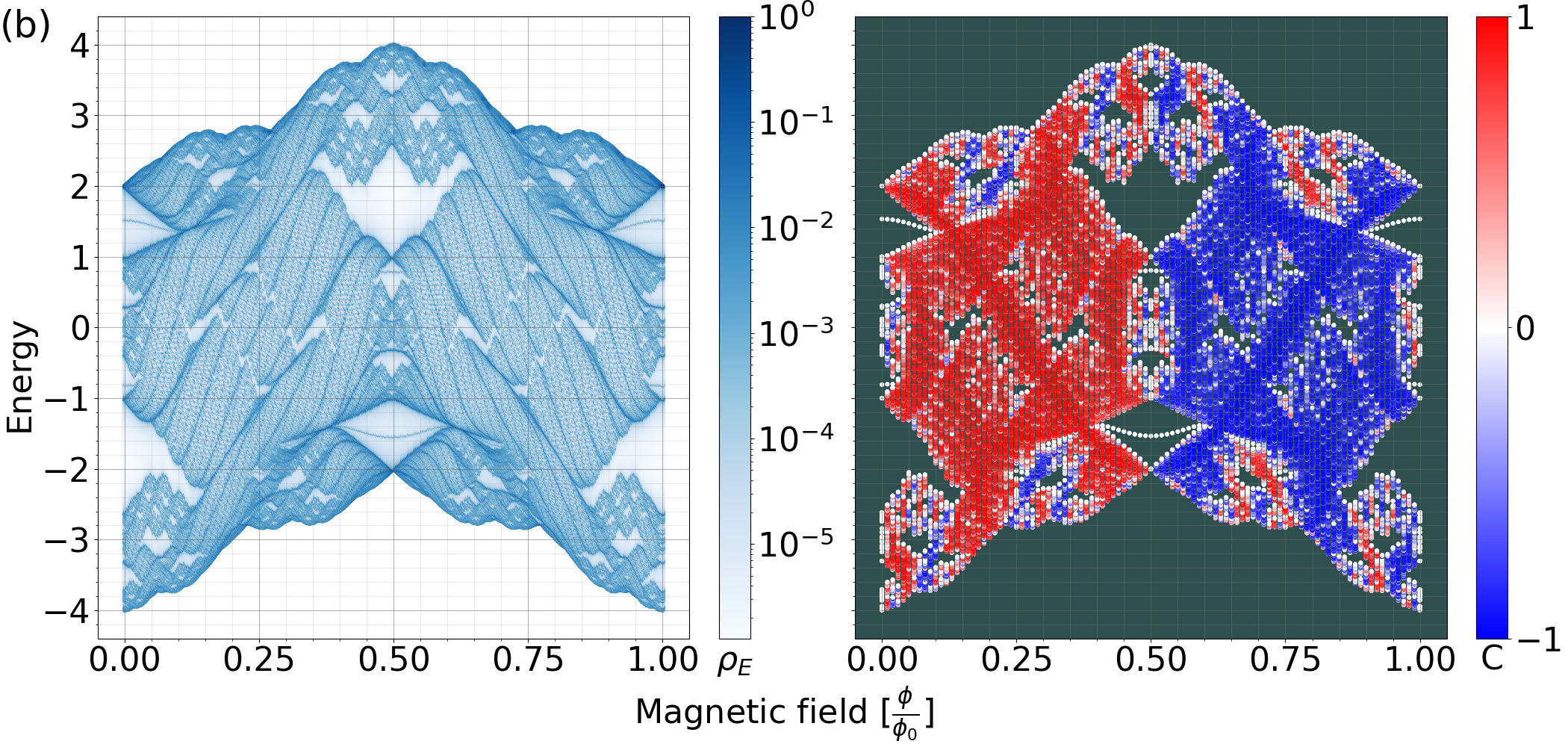}
\caption{Density of states and the real space Chern number for (a) SG-3, $N=3^6$ and, (b) SG-4, $N=(3^7+3)/2$. The density of states, $\rho_E$, is computed using Eq.\ \eqref{DOS_E} with $\epsilon=10^{-3}$ and the Chern number is computed using Eq.\ \eqref{KitaevChernNumber}.}
\label{DOS_chern}
\end{figure}

To study the spectrum, we look at the normalized density of states $\rho_E$, which is given by
\begin{equation}
\rho_E(E)=\sum_{n}\dfrac{1}{N}\delta(E-E_{n})=\sum_{n}\dfrac{1}{N\pi} \lim_{\epsilon \to 0}\dfrac{{\epsilon}}{(E-E_n)^2+{\epsilon}^2} 
\label{DOS_E}
\end{equation}
where $N$ is the total number of eigenstates and $n$ denotes the index of each eigenstate. We also calculate the real space Chern number for different fillings using Kitaev's prescription given by
\begin{equation}
C(P)=12\pi{\text{i}}(\text{Tr}(\mathbb{A}P \mathbb{B}P \mathbb{C}P)-\text{Tr}(\mathbb{A} P \mathbb{C} P \mathbb{B}P))
\label{KitaevChernNumber}
\end{equation}
where $\mathbb{A},\mathbb{B},\mathbb{C}$ are the projections onto the three partitions shown in Fig.\ \ref{Syst_&_partition}, and $P=\sum_{n \in occ} \ket{n}\bra{n}$ is the projector onto the set of occupied eigenstates. The normalized density of states and the Chern numbers as a function of the magnetic field are shown in Fig.\ \ref{DOS_chern}. It is immediately clear from Fig.\ \ref{DOS_chern} that most of the spectrum has a very low $\rho_E$. This is significantly different from the Hofstadter butterfly on lattices with open boundary conditions which have well-defined bulk regions (high $\rho_E$) and edge regions (low $\rho_E$) in the spectrum. Moreover, for both SG-3 and SG-4, almost the entire region with low $\rho_E$ is characterized by $C=\pm 1$ and all states in these regions are edge-like states. A few examples of such edge-like states can be found in references \cite{Adhip2018,Brzezinska2018,Sarangi2021}. 

\section{Adiabatic charge pumping}\label{Flux_pumping}
We insert an infinitely long, thin solenoid through a given point $(x_0,y_0)$. The flux, $2\pi\varphi$, through the solenoid is then varied adiabatically from $0$ to $2\pi$. We are interested in studying the response of the system to the change in flux.  To study that, it is important to study the many-body ground state of the system. Since the flux is pumped adiabatically and the Hamiltonian is non-interacting in nature, the many-body ground state of the system at a given instant is  the Slater determinant of the occupied single-particle eigenstates of the instantaneous Hamiltonian, $H(\varphi)$, with a dynamical and a geometric phase factor. So we first take a look at the single-particle eigenstates and eigenvalues of $H(\varphi)$. For the rest of the numerics in the text, given a state $\ket{\psi}=\sum_j \psi_j \ket{\textbf{r}_j}$, the localization is shown by computing the normalized onsite density, $\rho_j=|\psi_j|^2/{\text{max(}|\psi_j|^2\text{)}}$.

\subsection{Instantaneous spectrum and spectral flow}\label{spec_flow}

\begin{figure}
\includegraphics[scale=0.225]{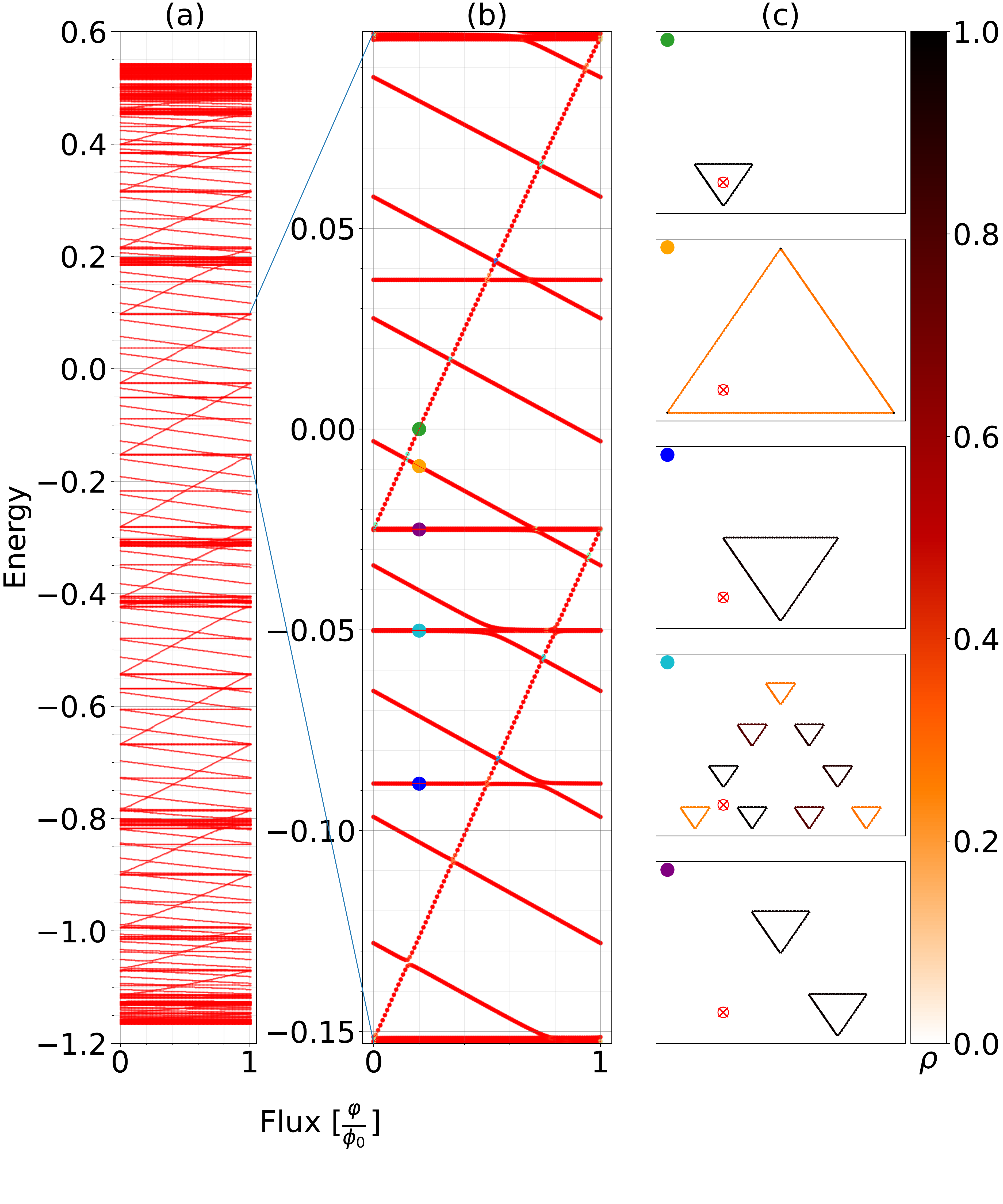}
\caption{Spectral flow for SG-3, $N=3^7$ with $\phi/{\phi_0}=0.3$. (a) shows the flow of the eigenstates as a function of $\varphi/{\phi_0}$ for a part of the spectrum. (b) is a zoomed in version of the spectral flow highlighting the nature of the flow. (c) shows the localization of different edge-like states corresponding to different spectral flow (computed at $\varphi/\phi_0=0.2$). The coloured dots in (b) represent the points at which the states in (c), marked with corresponding colors, were computed. The position of the flux tube is marked by a red cross-hair on the plots in  (c). }\label{Spectral_flow}
\end{figure}

The form of the Hamiltonian $H(\varphi)$ is the same as in Eq.\ \eqref{HH}, except that an additional Peierls phase, $\tilde{\theta}_{jk}=(e/h)\int_{\vec{r}_j}^{\vec{r}_k}\textbf{A}_{\varphi} \cdot \textbf{dl}$, gets added to each bond due to the flux-tube. Here $\textbf{A}_{\varphi}$ is the vector potential due to the flux tube, and for the numerical computations, it is taken to be $\textbf{A}_{\varphi}=(0, {\varphi}/r, 0)$ in cylindrical coordinates. The spectrum of $H(\varphi)$ at $\varphi=0$ and $\varphi=\phi_0$ are identical as the Hamiltonian returns to itself, up to a gauge transformation. In fact, the spectrum is periodic in $\varphi$ with a period of $\phi_0$. But for $\varphi \neq {n\phi_{0}}, {~~} n\in {\mathcal{Z}}$, the spectrum of the Hamiltonian changes in general resulting in the flow of the energy of individual eigenstates. We track the flow of the energies of the eigenstates as a function of $\varphi$. We say that a given state has undergone a spectral flow if the state does not return back to the same initial energy as $\varphi$ is changed from $0$ to $\phi_0$.
\newline

\begin{figure}
\includegraphics[scale=0.225]{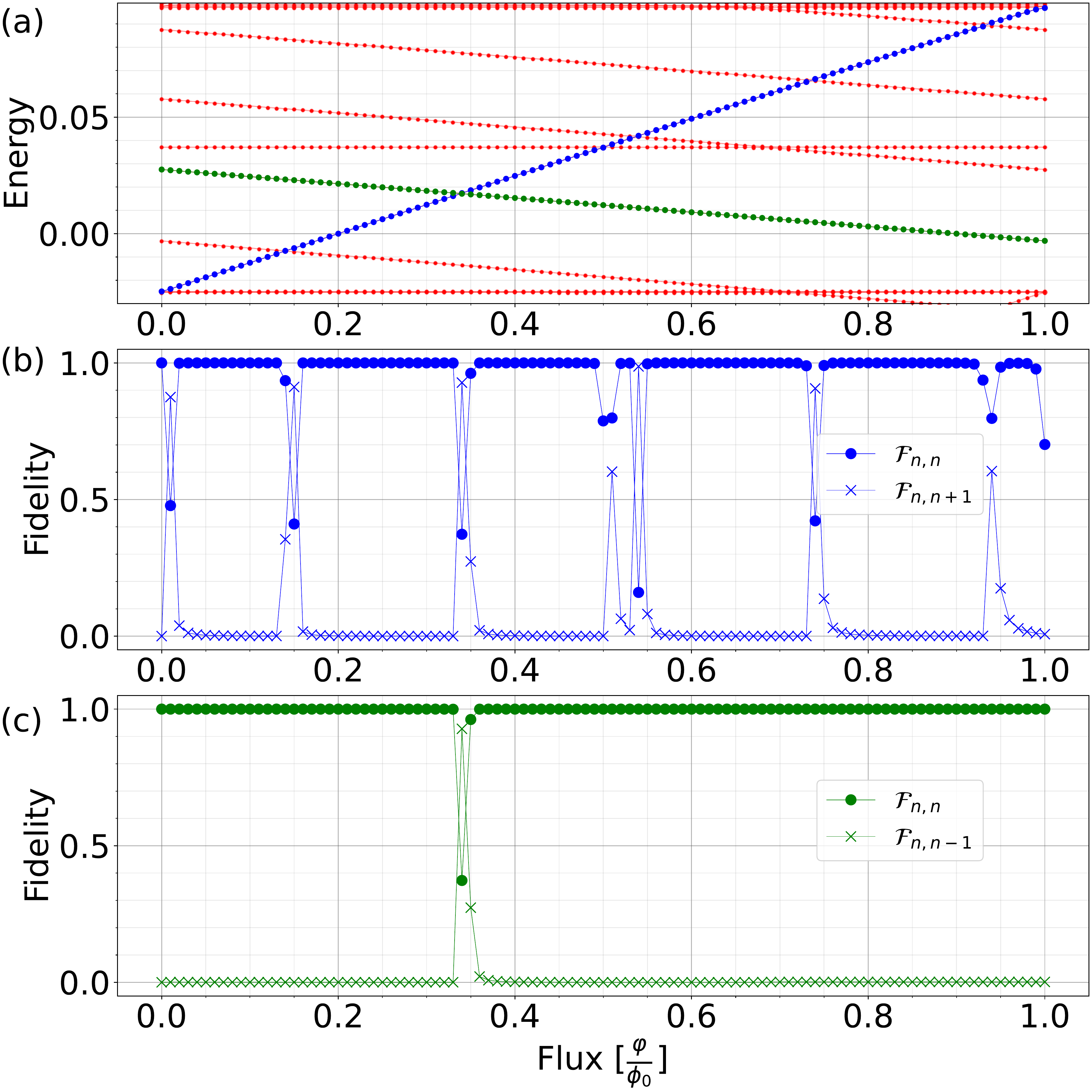}
\caption{Fidelity computations for avoided crossings. Fidelity is defined here as: $\mathcal{F}_{n,m}={\langle}{\psi_n}({\varphi})|{\psi_m({\varphi-{\delta \varphi}})}{\rangle}$. For the numerics, we have chosen $\delta \varphi=0.01\phi_0$ and $N=3^7$. (a) shows a part of the instantaneous spectrum when the flux tube is placed inside one of the triangles of the second generation (shown in Fig.\ \ref{Spectral_flow}(c)). The flow of two particular states, one of them being primarily localized on the sites immediately enclosing the triangle of the second generation and the other being localized on the sites of the outermost triangle of the SG, are marked in blue and green colors. (b) and (c) show the fidelity of these two states as a function of $\varphi$. Large dips in $\mathcal{F}_{n,n}$ and a correspondingly large peak in $\mathcal{F}_{n,n-1}$ ($\mathcal{F}_{n,n+1}$) are seen when the highlighted states come close in energy with another state localized on far off sites. These indicate that the state $\ket{\psi_n}$ has flowed to $\ket{\psi_{n-1}}$ ($\ket{\psi_{n+1}}$) without any significant hybridization. On the other hand, shallow dips in $\mathcal{F}_{n,n}$ and a correspondingly small peak in $\mathcal{F}_{n,n-1}$ ($\mathcal{F}_{n,n+1}$) are seen when the highlighted states come close in energy to a state localized on sites relatively close to the highlighted states. These indicate avoided crossings with significant hybridization. }\label{Fidelity_R_1}
\end{figure}

Figure \ref{Spectral_flow} shows the spectral flow of the Harper-Hofstadter model on SG-3. We would like to highlight the fact that the spectral flow here is qualitatively different from that of the Harper-Hofstadter model on a 2-dimensional lattice. In the case of a 2-dimensional lattice, spectral flow is observed across the band-gap. The states in the bulk undergoing spectral flow move in energy (up or down depending on the Chern number), from one band to the next band, across the gap. The edge states, which lie entirely in the gap, undergo spectral flow in the opposite direction to that of the bulk states. In contrast, in the case of SG-3, spectral flow is observed almost throughout the entire spectrum (in the low $\rho_E$ regions). Here, the states undergoing spectral flow go from one group of degenerate states with low degeneracy in the low $\rho_E$ region to another, as opposed to one `band' to another or one high $\rho_E$ region to another.
\newline

\begin{figure}
\includegraphics[scale=0.225]{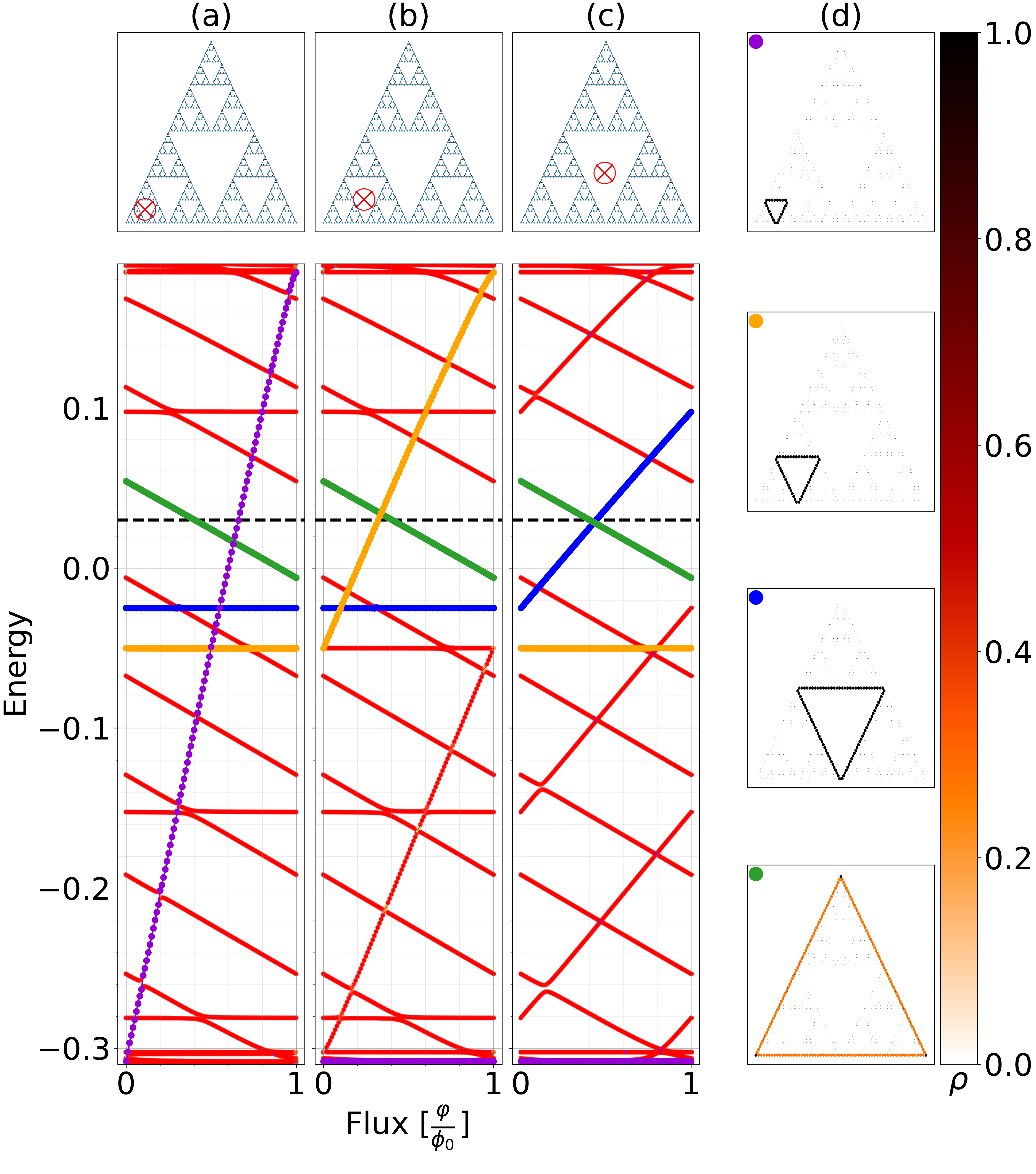}
\caption{Change in the spectral flow due to the change in the position of the flux tube for SG-3 ($N=3^6$, $\phi/{\phi_0}=0.3$). Columns (a), (b) and (c) show a portion of the spectral flow for three different positions of the flux tube. The position of the flux tube is marked by a red cross-hair in the SG-3 diagrams at the top of the  respective columns. A few typical states which are localized on sites enclosing holes of different generations are chosen and their spectral flows are highlighted with different colors. Column (d) shows the localization of these typical states ($\varphi/\phi_0=0.2$). To mark the correspondence, we have put circles of respective colors on the top-left corner of each of the localization plots. The black dashed line shows the position of the Fermi energy, $E_F=0.03$. The figure shows that only the edge-like states enclosing the flux tube undergo spectral flow.}\label{Flux_tube_var}
\end{figure}

The states in the low $\rho_E$ regions of the spectrum can be qualitatively grouped into four groups (column (b) of Fig.\ \ref{Spectral_flow}): (I) the states which flow up in energy (positive spectral flow), (II) states which flow down in energy (negative spectral flow), (III) states with almost no change in energy, but are degenerate at $\varphi=0$ with states undergoing spectral flow, and (IV) states with almost no change in energy and are not degenerate at $\varphi=0$ with states undergoing spectral flow. We find that the states in group I are edge-like states localized on the sites forming a loop which encloses the flux tube. The states in group II are edge-like states localized on the outermost triangle on SG-3. The states in group III and IV, which do not undergo a spectral flow, are also edge-like states but they are localized on sites forming loops which do not enclose the flux tube. We find that the real space localization is more or less the same for all states belonging to a given group, for low values of $\varphi$ and sufficiently away form the point of avoided crossings. This means that the states retain their edge-like localization away from the avoided crossings during the spectral flow. The representative real space localization of the states in the above mentioned groups, for a given position of the flux tube, are shown in column C of Fig.\ \ref{Spectral_flow} for these four groups. Close to the avoided crossings, the nature of the states changes due to hybridization. 
\newline

The extent of hybridization is dependent on the localization of the states; states localized nearby in real-space hybridize strongly in the absence of any symmetry. Here, the states belonging to different groups are edge-like states, localized on sites immediately enclosing triangles of different generations. Hence the extent of  hybridization is not significant. This has been checked by fidelity computations, shown in the Fig.\ \ref{Fidelity_R_1}. Fidelity is defined as: $\mathcal{F}_{n,m}={\langle}{\psi_n}({\varphi})|{\psi_m({\varphi-{\delta \varphi}})}{\rangle}$, where $\ket{\psi_n(\varphi)}$ and $\ket{\psi_m(\varphi)}$ are instantaneous eigenstates of the Hamiltonian ($H(\varphi)\ket{{\psi}_n(\varphi)}=E_n ({\varphi}) \ket{{\psi}_n(\varphi)} $), labeled by labels $n,m$ such that $n>m \implies E_n \geq E_m$. A high value of $\mathcal{F}_{n,m} \approx 1$ means that the state $\ket{\psi_m(\varphi- \delta \varphi)}$ flows to $\ket{\psi_n(\varphi)}$ without significant hybridization when the flux $\varphi$ is changed by an amount $\delta \varphi$.
\newline

Why certain states undergo spectral flow and certain states do not can be understood from their localization. For states belonging to group I and II, the states are always localized on a closed loop enclosing the flux tube. Hence, they are sensitive to flux (Aharonov-Bohm effect) and undergo spectral flow. On the other hand, the states which belong to group III and IV are localized on loops which do not enclose the flux tube. Hence, the vector potential of the flux tube can be effectively gauged out resulting in these states being not sensitive to the flux. As a result, they do not show spectral flow. This becomes further clear from Fig.\ \ref{Flux_tube_var} where we show the change in the spectral flow by changing the position of the flux. Clearly, a state localized on a given loop only undergoes spectral flow when the flux tube is enclosed within the loop. Also, for a bunch of degenerate edge-like states localized on different loops, the flux tube breaks the degeneracy if enclosed by one of the loops, resulting in spectral flow of only the state enclosing the flux tube (Fig.\ \ref{Loc_eig}).
\newline

There are a few other states in the spectra which we have not discussed in detail in this work. These states belong to the very few high $\rho_E$ regions in the spectra. In terms of localization, they are predominantly bulk-like in nature. Also, they do not show a clear spectral flow, owing to the high $\rho_E$ around them.

\subsection{Charge transport from the instantaneous spectrum}\label{charge_transport}
Let us consider a case where we have filled our system to a certain Fermi energy, $E_F$ (dashed black line in Fig.\ \ref{Flux_tube_var}). At $\varphi=n\phi_0:~n\in \mathcal{Z}$, let us denote the set of states with positive spectral flow (group I in Sec.\ \ref{spec_flow}) as $\lbrace\ket{{\psi^p}_m}\rbrace$ with energies $\lbrace E_m \rbrace$, and the set of states with negative spectral flow (group II in Sec.\ \ref{spec_flow}) as $\lbrace\ket{{\psi^n}_{m'}}\rbrace$ with energies $\lbrace E_{m'} \rbrace$. Here, $m$, $m'$ are the labels for the eigenstates localized on sites immediately enclosing a single triangle of a given generation of SG-3, such that their energies are ordered increasingly ($E_m<E_{m+1}$ for all $m$). Now, let us assume $E_F$ is such that $E_{m}<E_F<E_{m+1}$ and $E_{m'-1}<E_F<E_{m'}$ for some $m$, $m'$.
\newline

When we vary the flux adiabatically by a unit through the flux tube, the Hamiltonian returns back to itself (up to a gauge transformation), but the states undergoing spectral flow do not return back to themselves. In the beginning of the pumping cycle, ${\psi^p}_{m}$, ${\psi^n}_{m'-1}$ were occupied and ${\psi^p}_{m+1}$, ${\psi^n}_{m'}$ were empty. During the pumping cycle, the filled state ${\psi^p}_{m}$ gets pushed up in energy across $E_F$ and flows to ${\psi^p}_{m+1}$, and the empty state ${\psi^n}_{m'}$ flows down in energy across the $E_F$ to ${\psi^n}_{m'-1}$. As a result of this spectral flow, at the end of the pumping cycle, ${\psi^p}_{m+1}$ is filled and  ${\psi^n}_{m'-1}$ is empty. This spectral flow is observed for all $m$, $m'$ such that $E_{m}$ and $E_{m'}$ are away from the gaps (regions with zero $\rho_E$) in the energy spectrum at $\varphi=0$. And as long as $E_F$ is away from these gaps, exactly one state with positive spectral flow and one state with negative spectral flow cross the Fermi energy during the pumping cycle. Now, as pointed out earlier in the previous subsection, ${\psi^n}_{m'} \forall {m'}$ are localized on the sites on the outermost triangle and ${\psi^p}_{m} \forall {m}$ are localized on the closest sites enclosing the flux tube. So, when a unit flux is pumped, a single state localized on the outer-most sites of SG-3 is emptied and a single state localized near the flux tube gets filled, effectively pumping a unit charge radially from the outer-most loop to the loop closest to the flux tube. The mathematical details corresponding to the above arguments can be found in appendix\ \ref{adiabatic_transport}.
\newline

We want to highlight the local nature of the radial charge transport happening in this case. From the instantaneous spectrum (columns (a), (b), (c) of Fig.\ \ref{Flux_tube_var}), it is clear that edge-like states, localized on sites immediately enclosing different triangles of SG-3, undergo spectral flow and cross the Fermi energy as the position of the flux tube is changed. As described in the previous paragraph, only these states which flow across the Fermi energy contribute to the radial charge transport as a result of adiabatic pumping. So, given the position of the flux-tube and the Fermi energy, it is possible to exactly determine which edge-like states are contributing to the transport. Also, the position of the flux-tube can be used as a tuning parameter to selectively pump particles from sites immediately enclosing a particular triangle to the outermost triangle. In Sec.\ \ref{sec_hall_cond}, we also compute the local Hall conductivity in a slightly different setting which also reveals the local nature of the transverse charge transport in greater detail.
 
\section{Properties of Edge-like states}\label{prop_edge_states}
\begin{figure}
\includegraphics[scale=0.225]{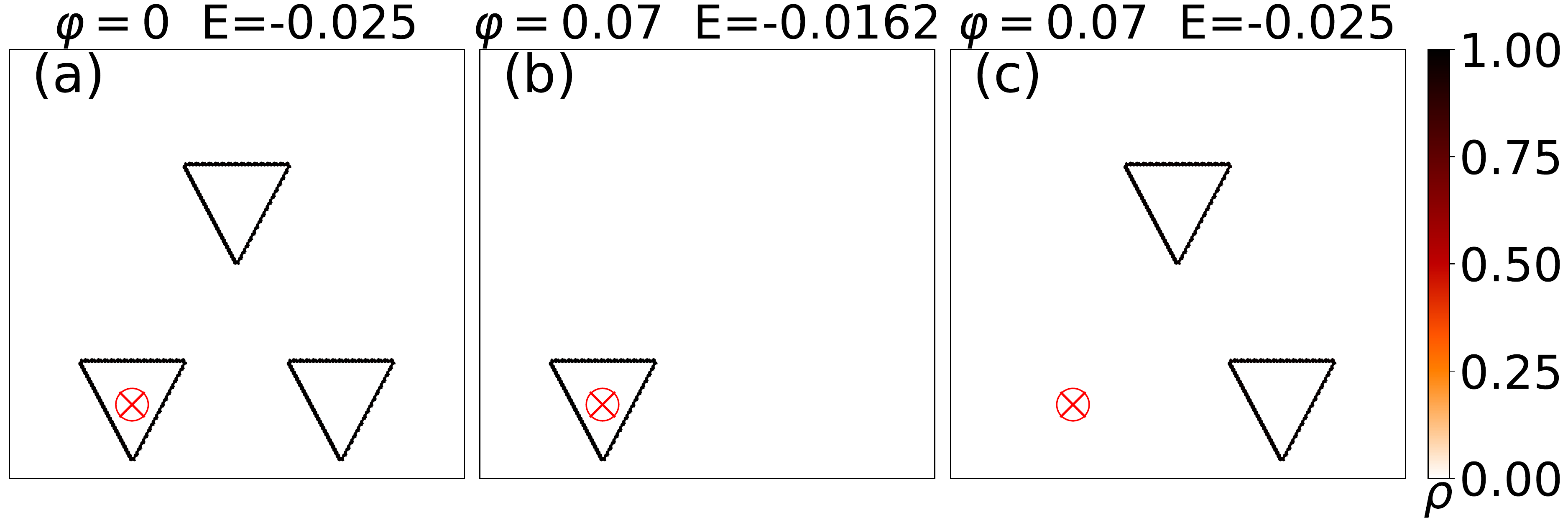}
\caption{The figure shows the degeneracy between edge-like states of SG-3  ($N=3^7$) at $\phi=0.3\phi_0$, highly localized on sites immediately enclosing triangles of the second generation, being lifted when the flux through the tube, $\varphi$, is changed from $0$ to $0.07\phi_0$. (a) shows the localization of a single eigenstate from a bunch of triply degenerate edge-like states at $\varphi=0$. Upon increasing $\varphi$ to $\varphi=0.07\phi_0$, the degeneracy breaks. One of the states, shown in (b), is lifted up in energy. The other two remain at the same energy as $\varphi=0$, one of which is shown in (c).} \label{Loc_eig}
\end{figure}

Pumping flux through a flux-tube at a given position not only makes it possible to determine how the states contribute to the transport, but it also makes it possible to numerically study each edge-like state individually. The edge-like state localized on sites immediately enclosing triangles of a given generation are usually degenerate in energy as there are often multiple triangles of a given generation in SG-3. One example is shown in Fig.\ \ref{Loc_eig}(a), where three states are degenerate, because there are three triangles of the second generation. The number of triangles of a given generation increases exponentially with the generation. So, it becomes hard to isolate a single edge-like state localized on the sites immediately enclosing a single triangle of high enough generation. Now, by positioning the flux tube in a given triangle, the energy of the edge-like state localized on sites enclosing that particular triangle increases as we pump flux through the tube (grouped into group I in Sec.\ \ref{spec_flow}). The energy of the remaining degenerate partners of that edge-like state does not change with flux as they localize on sites which do not enclose the flux-tube (grouped into group IV in Sec.\ \ref{spec_flow}). One such instance of degeneracy breaking is shown in Fig.\ \ref{Loc_eig}(b,c).
\newline 

\begin{figure}
\includegraphics[scale=0.225]{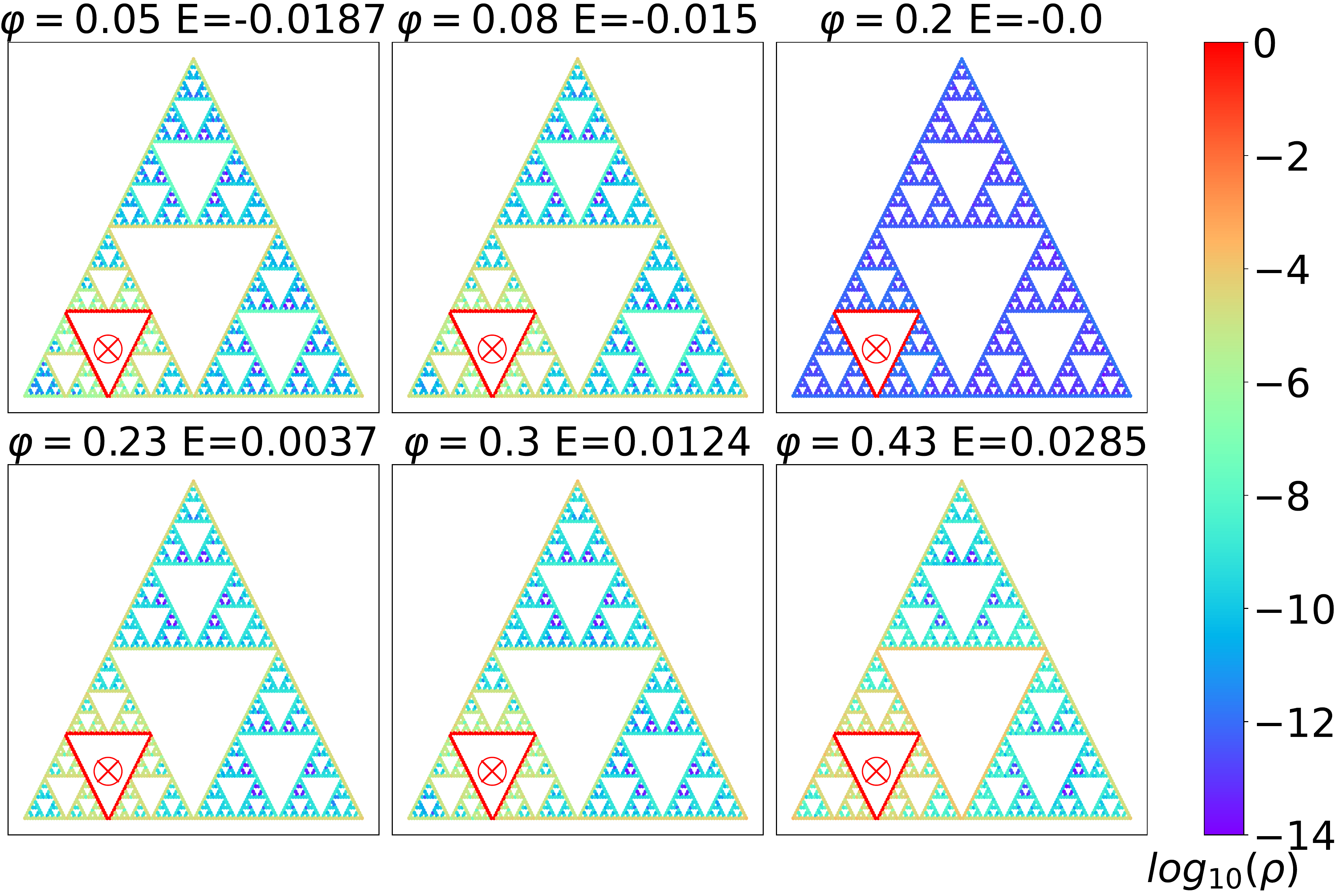}
\caption{Localization of a single edge-like state at different values of $\varphi$. The colorbar shows $\text{log}_{10}(\rho_j)$. The values of $\varphi$ in the plots are given in units of $\phi_0$ and the values of the $\varphi$s for this figure are chosen as such to remain significantly away from the avoided crossing points. We see that the weight of the eigenstate on sites not in $\mathcal{P}$ (defined in Sec.\ \ref{prop_edge_states}) is at least 3 orders of magnitude less than that of the sites in $\mathcal{P}$. $\varphi=0.2\phi_0$ is a special case where the edge-like state is completely localized on the sites in $\mathcal{P}$.} \label{Loc_phi_eig}
\end{figure}

Now that we are able to break the degeneracy, we can study the properties of a single edge-like state. An edge-like state on SG-3, by definition, is highly localized on the sites immediately enclosing a triangle of a given generation (Fig.\ \ref{Loc_phi_eig}). Let us denote the set of such sites by $\mathcal{P}$. Notice that the sites in $\mathcal{P}$, together with the bonds with their respective nearest neighbors in $\mathcal{P}$, form a ring and hence they can be indexed linearly from 1 to $N_{\mathcal{P}}$, where $N_{\mathcal{P}}$ is the total number of sites in $\mathcal{P}$. Now, given an edge-like state, $\ket{\psi^p_m}=\sum_j \psi^m_j \ket{\textbf{r}_j}$, we construct a state $\ket{\psi_m}=\sum_{\textbf{r}_j \in \mathcal{P}} \psi^m_j \ket{\textbf{r}_j}$. $\ket{\psi_m}$ is easy to study due to its one dimensional nature and can be considered a good approximation for $\ket{\psi^p_m}$ for large system sizes. 
\newline

\begin{figure}
\includegraphics[scale=0.225]{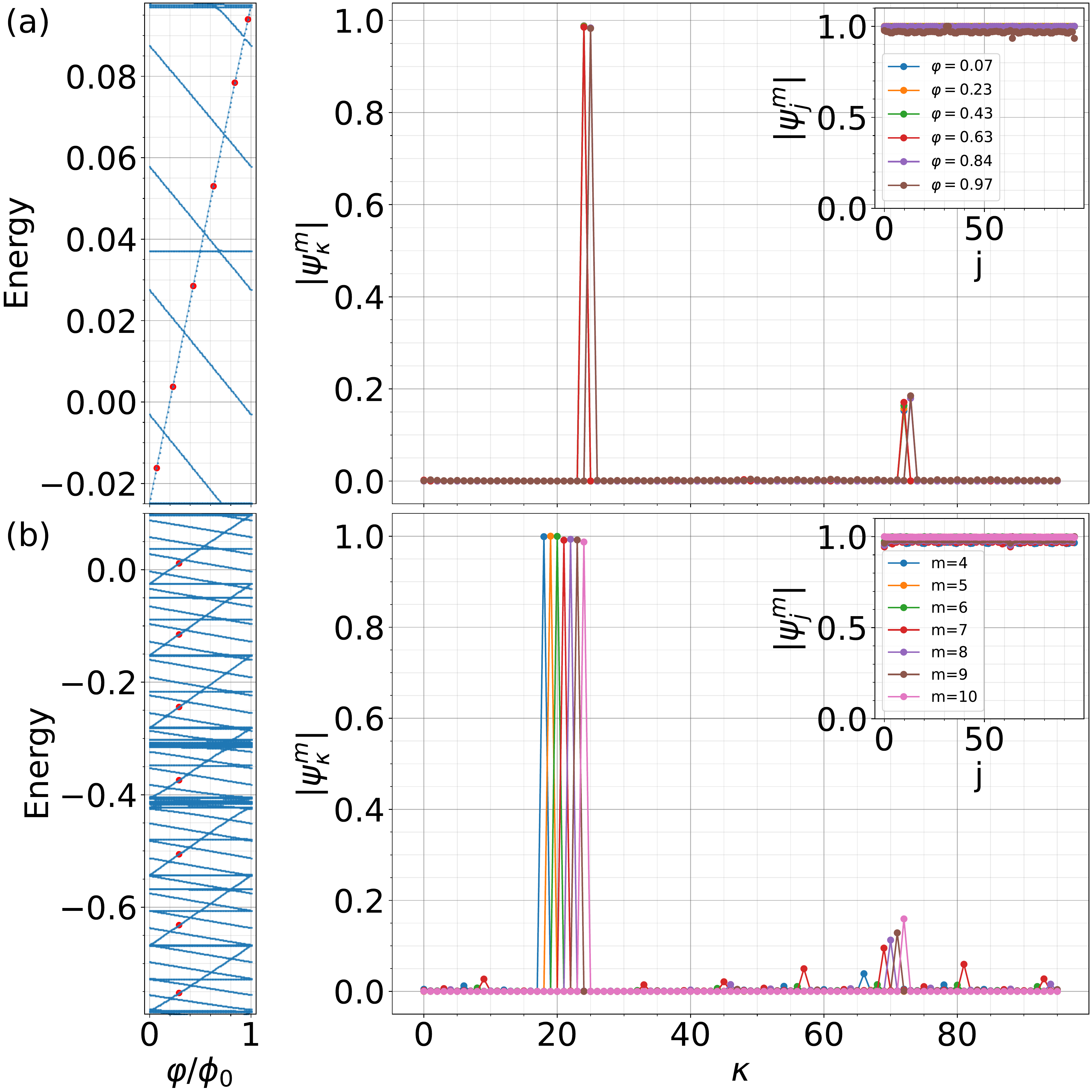}
\caption{The variation in Fourier amplitudes of edge-like states, $\psi^m_{\kappa}$ vs ${\kappa}$, for different values of $\varphi$ and different values of $m$. In this case, $N=3^7$ and $N_{\mathcal{P}}=96$. Row (a) shows $\psi^m_{\kappa}$ vs ${\kappa}$ for a single edge-like state ($m=10$). As $\varphi/\phi_0$ is varied from $0$ to $1$, the peaks at $\kappa=24$ and $N_{\mathcal{P}}/2+\kappa=72$ shift by one unit to $\kappa=25$ and $N_{\mathcal{P}}/2+\kappa=73$. Row (b) shows $\psi^m_{\kappa}$ vs ${\kappa}$ for different edge-like states which are primarily localized on the sites in $\mathcal{P}$. The insets in both (a) and (b) show how strongly the states are localized on the sites in $\mathcal{P}$. The states whose Fourier components are shown are marked with red on the instantaneous spectra. In both (a) and (b), the values of $\psi^m_{\kappa}$ have been normalized such that $\sum_{\kappa} {|\psi^m_{\kappa}|}^2 = 1$. } \label{Eig_prop}
\end{figure}

\begin{figure}
\includegraphics[scale=0.212]{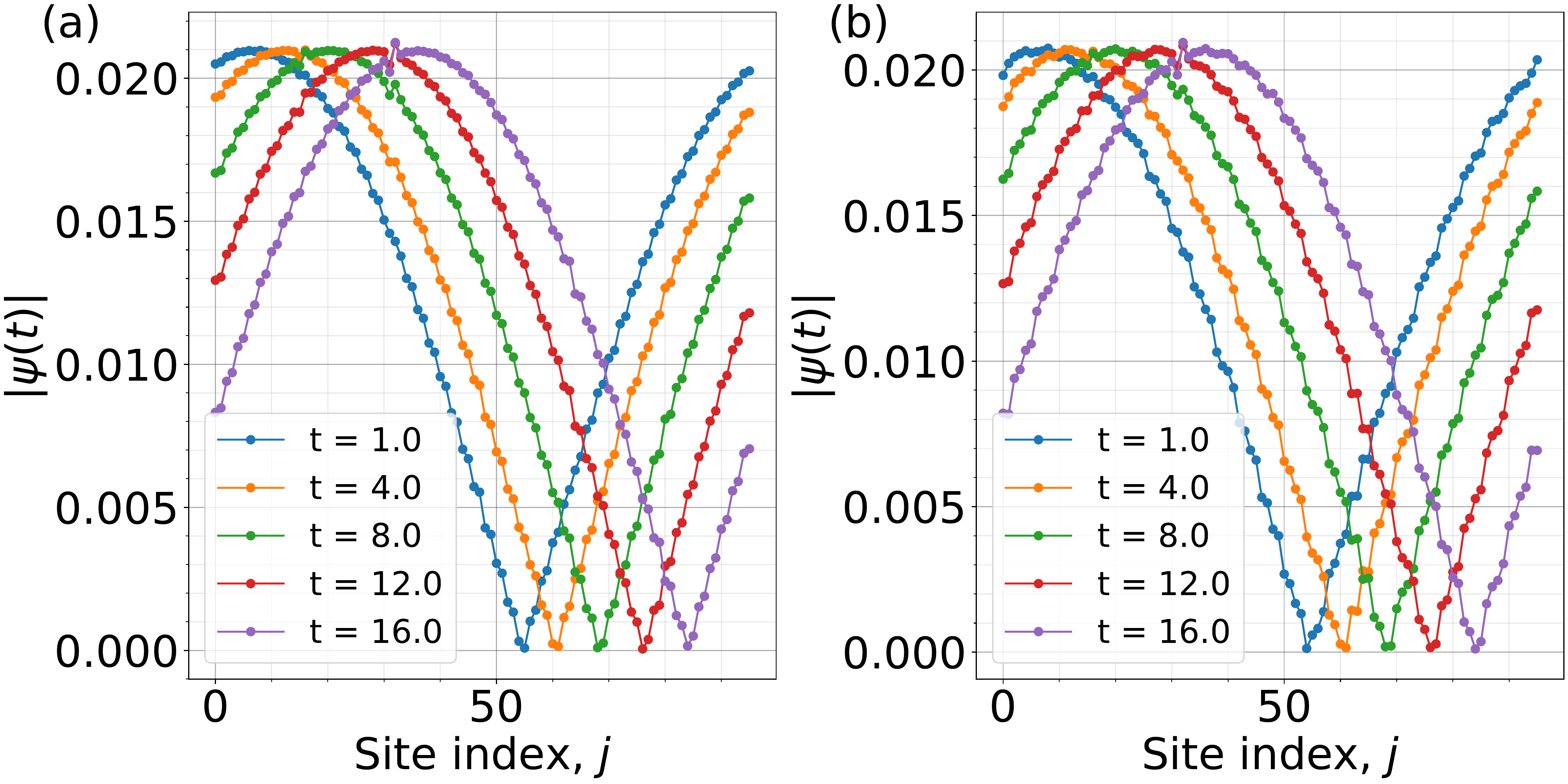}
\caption{Time evolution under the action of (a) the approximate time evolution operator and (b) the exact time evolution operator, projected onto a given energy window ($E_{\text{min}}, E_{\text{max}}$). The initial state is localized on a site in  $\mathcal{P}$. The exact time evolution operator, projected onto the energy window, is given by $U(t)=\sum_{E_{\text{min}} < E_n < E_{\text{max}}} \exp(-iE_n t ) \ket{n}\bra{n}$, where $\ket{n}$ is the set of single particle eigenstates of the Hamiltonian $H$, with energy $E_n$. The approximate time evolution operator is given by $\tilde{U}(t)=\sum_{E_{\text{min}} < \tilde{E}_n < E_{\text{max}}} \exp(-i\tilde{E}_n t ) \ket{\tilde{n}}\bra{\tilde{n}}$. $\ket{\tilde{n}}=\ket{\tilde{\psi}_n}$ for edge-like states localized primarily on $\mathcal{P}$, where $\tilde{\psi}_n$ is the approximation of the state using its first largest two Fourier components as mentioned in Sec.\ \ref{prop_edge_states}, and $\ket{\tilde{n}}=\ket{n}$ otherwise. $\tilde{E}_n = \ev{H}{\tilde{n}}$. Note that, at $t=0$, both $U$ and $\tilde{U}$ act as a projection operator onto the set of  states in the energy window. Comparing (a) and (b), we find that the chiral nature of the edge-like states is well captured by the approximate states mentioned in Sec.\ \ref{prop_edge_states}. For this calculation, we have taken $N=3^7$,  $N_{\mathcal{P}}=96$, $E_{\text{min}} = -0.3$ and $E_{\text{max}}=0$.   } \label{Time_evolution}
\end{figure}

The exact expression of the edge-like states is gauge dependent. So to study $\ket{\psi_m}$, we first transform to a different gauge where the Hamiltonian, $H$, becomes translationally invariant on the sites in $\mathcal{P}$. The gauge transformation is given by: $c^{\dagger}_j \rightarrow e^{-i\Theta_j}c^{\dagger}_j$, where $\Theta_1=0$, $\Theta_j=\sum^{n=j}_{n=2} \theta_{n-1,n} - (j-1)2\pi\Phi/N_{\mathcal{P}} \phi_0$ for $j \in \{2,3,\ldots,N_{\mathcal{P}}\}$, where $2\pi\Phi$ is the total flux threaded through the area enclosed by the sites in $\mathcal{P}$. Let us call this the `translationally invariant' gauge. Let the transformed state be denoted by $\ket{\psi'_m}=\sum_j \psi'^m_j \ket{\textbf{r}_j}$. We do a Fourier transform, $\psi^m_{\kappa}=\sum_j e^{i2{\pi}{\kappa}j/N_{\mathcal{P}}} \psi'^m_j$, to go into the angular momentum basis. We find that, for a given $m$, $\psi^m_{\kappa}$ has two sharp peaks at ${\kappa}$ and $N_{\mathcal{P}}/2 +{\kappa}$, (one peak being significantly greater than the other) for some value of ${\kappa}={\kappa}_0$ (Fig.\ \ref{Eig_prop}), as long as we are sufficiently away from an avoided crossing. The peaks change from ${\kappa}_0 \rightarrow {\kappa}_0+1$ and $N_{\mathcal{P}}/2+{\kappa}_0 \rightarrow N_{\mathcal{P}}/2+{\kappa}_0+1$ as the flux, $\varphi/\phi_0$, is changed from $0$ to $1$ (Fig.\ \ref{Eig_prop} (a)). Moreover, we also find that the position of the peaks changes linearly as we change $m$ (Fig.\ \ref{Eig_prop} (b)). These features are reminiscent of eigenstates of a particle on an $N_{\mathcal{P}}$-polygon with a flux threaded through it, or in other words, eigenstates of the discretized angular momentum operator \cite{Analytis2004}. In fact, these properties are captured by approximating, $\ket{\psi'_m} \approx \ket{\tilde{\psi}_m}=\psi^m_{\kappa_0}\ket{\kappa_0}+\psi^m_{\kappa_0+N_{\mathcal{P}}/2}\ket{\kappa_0+N_{\mathcal{P}}/2}$, where $\ket{\kappa_0}$ and $\ket{\kappa_0+N_{\mathcal{P}}/2}$ are eigenstates of the discretized angular momentum operator with eigenvalues $\kappa_0$ and $\kappa_0+N_{\mathcal{P}}/2$ respectively. $\ket{\tilde{\psi}_m}$ also captures the chiral nature of the edge-like states as shown in Fig.\ \ref{Time_evolution}.
\newline

For a given magnetic field parameterized by $\phi$, there are some special values of the flux $\varphi$ for which an edge-like state can be completely localized on the sites in $\mathcal{P}$ (for example, $\varphi=0.2\phi_0$ in Fig.\ \ref{Loc_phi_eig}). For such values of $\varphi$, the edge-like state exactly becomes an eigenstate of the discretized angular momentum operator in the `translationally invariant' gauge. This shows that the flux through the flux-tube can also be used as a tuning parameter to completely localize an edge-like state on a ring and host exact eigenstates of the angular momentum operator on SG-3. The details of the condition which must be satisfied to generate such an edge-like state is given in appendix \ref{Complete_localization_condition}.
\newline

\section{Local Hall conductivity and its robustness to disorder }\label{sec_hall_cond}
In this section, we study local contributions to the Hall conductivity, following the approach of \cite{Peru2022}. Specifically, we look at the Hall response of the system when the system is subjected to a step-function electric potential. To do this, we consider a horizontal cross-section at some $y=y_0$ and raise the potential of the system below this cross-section by $-V_0$. Such a potential can be treated in a time dependent gauge, $\textbf{A(t)}=(0, -A(t) \delta (y-y_0), 0)$, where $A(t)=V_0 t$. The time dependent Hamiltonian then  becomes $H(t)=e^{iA(t)\vartheta(y_0)}He^{-iA(t)\vartheta(y_0)}$, where $\vartheta(y_0)=\sum_j \theta(y_j-y_0)\ket{\textbf{r}_j}\bra{\textbf{r}_j}$. Now, working in the adiabatic limit, we look at the transverse current across a vertical cross-section at some $x=x_0$. For a non-interacting finite system in the above mentioned setting, it has been shown that the site-resolved Hall-conductivity, $\sigma_{xy}(r)$, can be expressed as a local Chern marker in the adiabatic limit \cite{Peru2022}. We highlight here the main ideas leading to this result in the context of our system. The details of the calculation can be found in Ref.\ \citep{Peru2022} and references therein.
\newline

\begin{figure}
\includegraphics[scale=0.2]{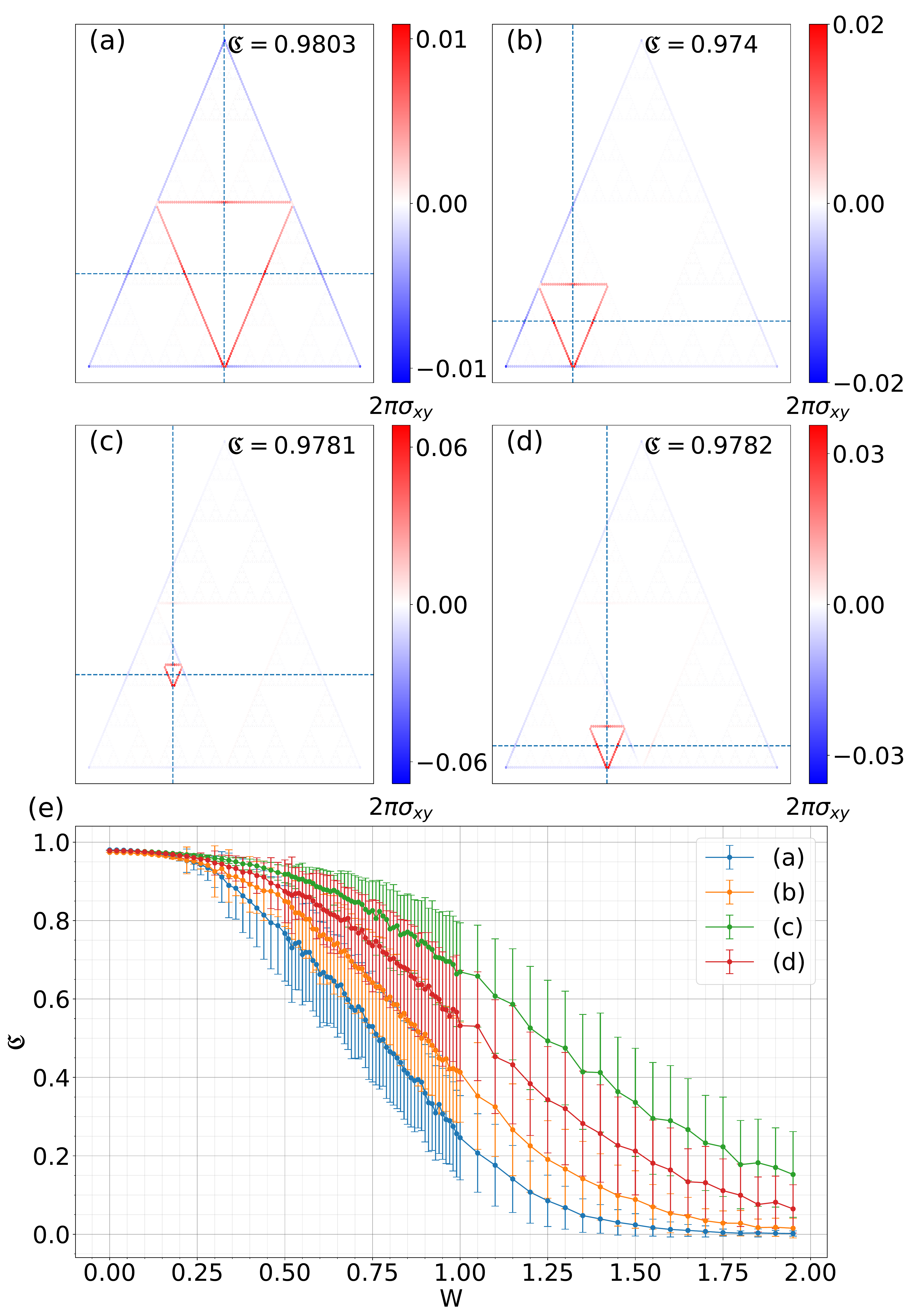}
\caption{(a-d) Site-resolved Hall conductivity, $\sigma_{xy}$,  for the Harper-Hofstadter model on SG-3 with $N=3^7$, for $E_F=0.03$. The dashed horizontal line represents the cross-section across which the potential difference is applied. The dashed vertical line represents the cross-section across which the current has been calculated. The sum over local Hall conductivity, $\mathfrak{C}=\sum_{r \in \mathcal{P}}2{\pi}\sigma_{xy}(r)$, where $\mathcal{P}$ denotes the set of sites which immediately enclose the cross-hair, are mentioned on the plots up to four decimal places. (e) Variation of the local sum of Hall conductivity, $\mathfrak{C}$, as a function of disorder strength $W$ for different positions $R$ of the cross-hair as in (a-d), calculated for the Harper-Hofstadter model with onsite Anderson disorder on SG-3 with $N=3^7$, for $E_F=0.03$. Averaging has been done with, $N_w=200$, disorder realizations. The error bars on the plot show the statistical standard deviation of $\mathfrak{C}$ over the disorder realizations. }
\label{cond_dis_var}
\end{figure}

Given that we are interested in the adiabatic limit, we use the adiabatic Hamiltonian, $K(t)=i[\dot{P}_I,P_I]$, to generate the time-evolution. Here, $P_I$ is the instantaneous projection operator onto the occupied states defined as $P_I=\sum_{E_n<E_F}\ket{n(t)}\bra{n(t)}$, where $H(t)\ket{n(t)}=E_n\ket{n(t)}$. With the adiabatic Hamiltonian, the instantaneous projection operator satisfies the von Neumann equation
\begin{equation}
\partial_tP_I(t)=-i[K,P_I].
\end{equation}
Given the form of $H(t)$, it is clear that $P_I=e^{iV_0t\vartheta(y_0)}Pe^{-iV_0t\vartheta(y_0)}$, and hence $K(t)=\dot{A}(P\vartheta(y_0)Q+Q\vartheta(y_0)P)$, where $P=P_I(t=0)$ and $Q=\mathbb{1}-P$.
The adiabatic transverse current operator, $J_x(t)$, can be obtained from the rate of change of the number of particles present in one side of the vertical cross-section using the instantaneous von Neumann equation, as $J_x(t)=i[K(t),\vartheta(x_0)]$, where $\vartheta(x_0)=\sum_j \theta(x_j-x_0)\ket{\textbf{r}_j}\bra{\textbf{r}_j}$. The site-resolved current operator can be defined as 
\begin{equation}\label{loc_current}
J_x^A(\textbf{r}_j,t)=\dfrac{1}{2}\{\delta_j,J_x^A(t)\}, 
\end{equation}
where $\delta_j=\ket{\textbf{r}_j}\bra{\textbf{r}_j}$. So, the site-resolved adiabatic current, $\langle J_x^A(\textbf{r}_j,t)\rangle=\text{Tr}_{\textbf{r}_j}(P_I J_x^A(\textbf{r}_j,t))$, is then given by
\begin{equation}
\langle J_x^A(\textbf{r}_j,t)\rangle=i \dot{A}\text{Tr}_{\textbf{r}_j}(P\vartheta(x_0)Q\vartheta(y_0)P)+h.c..
\end{equation}
Now, identifying $\dot{A}=-E$, we get the expression for Hall conductivity, $\sigma_{xy}={\langle J_x^A(\textbf{r}_j,t)\rangle}/E$, as
\begin{equation}
\sigma_{xy}({\textbf{r}_j})=2\text{Im}\text{Tr}_{\textbf{r}_j}(P\vartheta(x_0)Q\vartheta(y_0)P).
\end{equation}
\newline

The local Chern marker is then  defined as $\mathfrak{C}({\textbf{r}_j})=2\pi\sigma_{xy}({\textbf{r}_j})$. This has been referred to as the cross-hair marker in Ref.\ \cite{Peru2022} due to the fact that the horizontal line at $y=y_0$ and $x=x_0$ appear as a cross-hair. An important thing to note here is that the local Chern marker defined above is not unique for a given system as the way to define the site-resolved adiabatic current is not unique. The definition given in Eq.\ \eqref{loc_current} is  one simple way to define such a local quantity. Instead, the quantity which is physically relevant is the sum of the local Chern marker over some given region. This is because the sum of the local Chern marker over a region can be expressed as the Hall conductivity which derived from the total current leaking from that given region. The current leaking from that given region is defined as the rate of change of particles over the region and does not have ambiguity in its definition as opposed to the site-resolved adiabatic current.
\newline

Figure \ref{cond_dis_var} (a-d) shows the adiabatic site-resolved Hall conductivity calculated for our system. For the purpose of comparing to the charge pumping picture, we have kept the same Fermi energy for these computations as that for the charge pumping computations. We find that there are two significant local contributions to $\sigma_{xy}$, one positive and one negative, as $\sum_{j} \sigma_{xy}({\textbf{r}_j})=0$ due to the conservation of particle number over the entire system. Fixing the cross-section ($y=y_0$) across which the potential difference is applied, when we change the cross-section ($x=x_0$) across which the transverse current is calculated, we find numerically that the positive contributions to $\sigma_{xy}$ come only from the sites close to the position of the cross-hair. More specifically, we find that, given a position of the cross-hair, the positive contributions to $\sigma_{xy}$ come only from the sites which enclose the cross-hair as long as the cross-hair is not inside one of the smallest triangles of the structure. The negative contribution comes solely from the sites on the outer-most triangle of SG-3. What this suggests is that, as long as we are away from the smallest possible triangles of the structure, the contribution to the transverse current comes primarily from the sites which immediately enclose the cross-hair, or in other terms, from the edge-like states which are localized on the sites enclosing the triangle containing the cross-hair. This shows the correspondence between the local contribution to the Hall conductivity and the  local nature of the transverse charge transport in this system mentioned in Sec.\ \ref{charge_transport}. 
\newline

To see the quantized nature of the charge transport, we look at the sum of the local Chern marker over the sites in the proximity of the cross-hair. More specifically, we look at
\begin{equation}
\mathfrak{C}=\sum_{{\textbf{r}_j} \in \mathcal{P}} 2\pi \sigma_{xy}({\textbf{r}_j}),
\end{equation}
where $\mathcal{P}$ denotes the set of sites which immediately enclose the smallest triangle containing the cross-hair. We consider this quantity as it is physically relevant and it tells about the net charge leaking from the region containing the sites in $\mathcal{P}$. This can be expressed as the change in the projector over the occupied states, traced over the given region when the Hamiltonian is taken in a cycle \cite{Peru2022, Kitaev2006, Avron1994}. Hence, this quantity would be quantized if the change in the projector has support only in the region we trace over. In this system, we find that the value of $\mathfrak{C}$ is closely quantized to $1$, as mentioned in the plots in Fig.\ \ref{cond_dis_var} (a-d), suggesting again that significant contribution to the radial current comes from the edge-like states localized on sites in $\mathcal{P}$. 
\newline

To see the topological nature of the charge transport, we perturb the Hamiltonian slightly by adding small on-site disorder. The new disordered Hamiltonian is then given by
\begin{equation}
H_{dis}=H+\sum_j \epsilon_j c^{\dagger}_jc_j,
\end{equation}
where $\epsilon_j$ is a random number with a uniform distribution over the interval $[-W/2,W/2]$. For a given Fermi energy, we compute $\sigma_{xy}(\textbf{r}_j)$ for various disorder realizations of the same disorder strength $W$. We find that, for $W \neq 0$, the contribution to $\sigma_{xy}$ now not only comes from the sites in $\mathcal{P}$, but also spreads over to few other sites in the proximity of $\mathcal{P}$. This spread increases initially as we increase $W$ until the states become Anderson localized and $\sigma_{xy}(\textbf{r}_j)$ goes to zero. To quantify this spread and study the robustness to disorder, we then look at how $\mathfrak{C}$, averaged over several disorder realizations, changes as a function of disorder strength $W$. The result is shown in Fig.\ \ref{cond_dis_var} (e). 
\newline

We find that up to $W \approx 0.2$, the value of $\mathfrak{C}$ is pretty well quantized and robust to disorder. As we keep increasing $W$, the average value of $\mathfrak{C}$ starts decreasing and the standard deviation, shown as error bars in Fig.\ \ref{cond_dis_var} (e), starts increasing. The initial decrease in the average value of $\mathfrak{C}$ is a consequence of the  increase of the contribution to $\sigma_{xy}$ coming from the sites not present in $\mathcal{P}$. The standard deviation can be considered as an indicator of the amount of variation of the contribution to $\sigma_{xy}$ is coming from the sites not present in $\mathcal{P}$, which are found to be random in nature. This spread can be understood by the fact that in the presence of weak disorder, the edge-like states start to lose their property of being primarily localized on the sites in $\mathcal{P}$. It is natural to ask if there is a better quantization at higher $W$, by redefining $\mathfrak{C}$ to take into account the contributions of a few additional layers of sites apart form those in $\mathcal{P}$ to $\sigma_{xy}$. However, for this structure, there is no natural way to determine how to select sites to define a layer of sites and how many additional layers of sites to take into account. Also, because of the non-uniformity in the spread of the edge-like states to the nearby sites in the presence of weak random disorder, it is not clear how to determine a length scale by quantifying their loss of localization.

\section{Summary and outlook}\label{summary}
In this article, we have studied the adiabatic charge pumping and transport of non-interacting fermions on self-similar structures generated from the Sierpinski gasket. We consider the Harper-Hofstadter Hamiltonian on SG-3 and SG-4, with an additional flux tube to adiabatically pump the charge. Since the systems are non-interacting and we are interested in the case where the pump works in the adiabatic limit, we study their respective instantaneous eigen-spectra. For SG-3, we find that, for a given position of the flux tube, all edge-like states  throughout the instantaneous spectrum, which are localized on sites enclosing the flux-tube, undergo spectral flow. This is qualitatively different from the spectral flow in the case of translationally invariant non-interacting systems where spectral flow is observed across the band gaps. Changing the position of the flux tube results in a change of the set of edge-like states undergoing spectral flow. We have found similar results for SG-4 which we have not shown here.
\newline

We find that the local nature of the adiabatic charge transport is also dependent on the position of the flux-tube. The transport happens between the sites hosting an edge-like state enclosing the flux-tube and the outermost sites of SG-3, which also host an edge-like state. However, the net charge transported is quantized, irrespective of the position of the flux-tube. We show that the quantization of the adiabatic charge and hence the topological character of the system, can be understood from the spectral flow occurring near the Fermi energy. Specifically, the adiabatic charge transported is non-trivially quantized if at least one pair of edge-like states, localized significantly far from each other, undergo opposite spectral flow crossing the Fermi energy.
\newline

We also  study the local Hall conductivity by explicitly computing the local transverse current when the system is subjected to a local step potential. We find that the local contributions to the Hall conductivity only comes from the sites which host the edge-like states enclosing the cross-hair, thus establishing a correspondence with the spectral flow of the edge-like states. We find that the total local contribution to the Hall conductivity is quantized and is robust to weak Anderson disorder. Upon increasing the disorder strength, the contribution to the local Hall conductivity does not remain highly localized anymore, before finally going to zero at high disorder strengths.
\newline

We use the flux in the flux-tube as a tuning parameter to isolate a single edge-like state from its degenerate group of states. We find that the edge-like states can be approximated by a sum of a few eigenstates of the discretized angular momentum operator. Our results suggest that instead of treating them to be analogous to the topological edge states in translationally invariant non-interacting systems, some of their properties can be understood from a perspective of a particle on a tight-binding polygonal chain with a non-zero flux. 
\newline

In conclusion, we have explained the microscopic origin of the topological character and the quantization of the Hall conductivity in self-similar structures, generated from the Sierpinski gasket, using the perspective of spectral flow and adiabatic charge pumping. We expect our results to generalize to a wider variety of self-similar structures and finite systems embedded in two dimensions, given that the systems are able to support eigenstates which are localized on sites which form loops in the graph of the Hamiltonian. More specifically, if a finite system, embedded in two dimensions, is able to support at least two different sets of eigenstates, localized on two different loops such that one loop completely encloses the other and are spatially separated from each other, then we expect the system to show spectral flow when the flux through the inner loop is varied adiabatically. And as a result, we expect such systems to show quantized Hall response. It is still unclear what kind of self-similar structures or finite systems in general, would support such states localized on loops. Also, among self-similar structures, every structure has an unique fundamental self-similar repeating unit which is iteratively used to generate the structure of higher generations. The relation between the structure of such fundamental self-similar repeating unit and the ability of the system to support localized states on loops is not known yet. These can be some potential directions for future work in this area.

\begin{acknowledgments}
We thank Aniket Patra, Hadi Yarloo and Blazej Jaworowski for useful discussions. This work has been supported by the Independent Research Fund Denmark under grant number 8049-00074B.
\end{acknowledgments}

\appendix
\section{Adiabatic charge transport in finite systems in terms of instantaneous projectors}\label{adiabatic_transport}
We consider a finite non-interacting system, $\mathcal{S}$, with a Hamiltonian, $H(\varphi (t))$, where $\varphi(t)$ is a time dependent parameter. We assume that the Hamiltonian has no other explicit time dependence and from now on, in this section, we suppress the time dependence of the parameter. The instantaneous eigenstates can be obtained from the eigenvalue equation
\begin{equation}
H(\varphi)\ket{n(\varphi)}=E_n(\varphi)\ket{n(\varphi)}.
\end{equation}
We assume that there is a time $t=T$ after which the Hamiltonian returns back to itself, up to a gauge transformation. We now consider a subsystem, $\mathcal{B}$, of the system. The rest of the system is denoted by $\mathcal{S}-\mathcal{B}$. We want to quantify the net charge, $Q$, leaking out of the subsystem over a time period when the system is adiabatically evolved in time. $Q$ is given by
\begin{equation}
Q=\int_0^T \ev{J}dt,
\end{equation}
where $J$ is the current operator and $\ev{}$ is the expectation value of the operator in the many-body ground state wavefunction at time $t$. In the adiabatic limit, each single particle eigenstate of the Hamiltonian $H(\varphi(t))$ evolves as
\begin{equation}
\ket{n(\varphi(t))}=e^{i\theta_n(t)}e^{i\gamma_n(t)}\ket{n(\varphi(0))},
\end{equation}
where $\theta_n(t)=-(1/{\hbar})\int_{0}^{t}E_{n}(\varphi(t'))dt'$ is the dynamical phase and $\gamma_n(t)=\int_{0}^{t}i\bra{n(\varphi(t'))}\ket{\dot{n}(\varphi(t'))}dt'$ is the geometrical phase. So the many-body time-evolved state in the adiabatic limit, $\ket{\Omega(t)}$, is  the Slater determinant of the adiabatically time-evolved occupied single particle states.
\newline

The current operator can be identified from the change of the total number operator over subsystem, $\mathcal{B}$, which is given by the von Neumann equation
\begin{equation}
\dfrac{\partial \ev{n_{\mathcal{B}}}}{\partial t}= -i\ev{[n_{\mathcal{B}},H(\varphi)]},
\end{equation}
where $n_{\mathcal{B}}=\sum_{b \in \mathcal{B}} \ket{\textbf{r}_b}\bra{\textbf{r}_b}$ is the total number operator over $\mathcal{B}$. Then we identify the current operator as $J=-i[n_{\mathcal{B}},H(\varphi)]$. In the case of adiabatic evolution, the time-evolution can be generated by the adiabatic Hamiltonian, $K(t)=i[\dot{P_I},P_I]$, instead of $H$ \cite{Peru2022}. Here $P_I=\sum_{n(\varphi) ~\in~ {occ}} \ket{n(\varphi)}\bra{n(\varphi)}$ is the instantaneous projector onto the set of occupied single particle states. The derivation of the adiabatic Hamiltonian, $K$, can be found in appendix A of Ref.\ \cite{Peru2022}. So, the adiabatic current operator is given by 
\begin{equation}
\begin{split}
J^A & =-i[n_{\mathcal{B}},K(t)] \\
& =(n_{\mathcal{B}} \dot{P_I} P_I + P_I \dot{P_I} n_{\mathcal{B}} - n_{\mathcal{B}} P_I \dot{P_I} - \dot{P_I} P_I n_{\mathcal{B}}).
\end{split}
\end{equation}

The expectation value of the adiabatic current operator in the many-body ground state then becomes
\begin{equation}
\begin{split}
\ev{J^A} & =\ev{J^A}{\Omega(t)}=\text{Tr}(P_I J^A)\\
&= \text{Tr} (P_I n_{\mathcal{B}} \dot{P_I} P_I) +\text{Tr} (P_I^2 \dot{P_I} n_{\mathcal{B}}) \\
&-\text{Tr} (P_I n_{\mathcal{B}} P_I \dot{P_I}) - \text{Tr}(P_I \dot{P_I} P_I n_{\mathcal{B}})\\
&= \text{Tr} (n_{\mathcal{B}} \dot{P_I} P_I) +\text{Tr} (P_I \dot{P_I} n_{\mathcal{B}}) \\
&-\text{Tr} (n_{\mathcal{B}} P_I \dot{P_I} P_I) - \text{Tr}(P_I \dot{P_I} P_I n_{\mathcal{B}}),
\end{split}
\end{equation}
where the last equality has been obtained by using the cyclic property of the trace and the fact that $P_I^2=P_I$. Now we use the identity $P_I \dot{P_I} P_I=0$, and we get
\begin{equation}
\begin{split}
\ev{J^A}&=\text{Tr} (n_{\mathcal{B}} \dot{P_I} P_I) +\text{Tr} (P_I \dot{P_I} n_{\mathcal{B}})\\
&=\text{Tr} ( n_{\mathcal{B}} \dot{P_I} P_I ) +\text{Tr} ( n_{\mathcal{B}} P_I \dot{P_I} )\\
&=\text{Tr} ( n_{\mathcal{B}} \dot{P_I^2})= \text{Tr}_{\mathcal{B}} (\dot{P_I}),
\end{split}
\end{equation}
where $\text{Tr}_{\mathcal{B}}(..)$ is the trace over degrees of freedom in subsystem $\mathcal{B}$. So the net charge leaking form $\mathcal{B}$ can be expressed as
\begin{equation}\label{Charge_leak}
\begin{split}
Q &=\int_0^T \ev{J^A} dt= \int_0^T \text{Tr}_{\mathcal{B}} (\partial_t P_I) dt\\
&= \text{Tr}_{\mathcal{B}} (P_I(T)-P_I(0)).
\end{split}
\end{equation}
\newline

From Eq.\ \eqref{Charge_leak} we see that, in the adiabatic limit, the net charge leaking from the subsystem can be expressed as the change in the instantaneous projector onto the occupied states over the pumping cycle, traced over the degrees of freedom of the subsystem. Now, as the parameter is assumed to get back to its initial value at time $T$, the Hamiltonian returns back to itself, up to a gauge transformation. So, the set of projectors onto the eigenstates of the Hamiltonian at $t=0$, $\lbrace\ket{n(\varphi(0))}\bra{n(\varphi(0))}\rbrace$, is the same as the set of projectors onto the eigenstates at $t=T$, $\lbrace\ket{n(\varphi(T))} \bra{n(\varphi(T))}\rbrace$. So, if there is no spectral flow due to the change in $\varphi$ in the instantaneous spectra of the Hamiltonian, $P_I(T)=P_I(0)$  and there is no adiabatic charge transport as a result. Clearly, to get a non-zero adiabatic charge transport from the region $\mathcal{B}$, there must be spectral flow in the instantaneous spectra of the system. 
\newline

Now let us consider a scenario where $P_I(T) \neq P_I(0)$. Let $\mathcal{N}_i$ be the set of eigenstates which are occupied at $t=0$ but not at $t=T$, $\mathcal{N}_f$ be the set of eigenstates which are occupied at $t=T$ but not at $t=0$, and $\mathcal{O}$ be the set of eigenstates which remain occupied both at $t=0$ and $t=T$. As we have assumed that the system is particle conserving, the number of states in $\mathcal{N}_i$ and $\mathcal{N}_f$ are the same, denoted by $N$. So, $P_I(0)=\sum_{n \in \mathcal{N}_i} \ket{n}\bra{n} + \sum_{o \in \mathcal{O}} \ket{o}\bra{o}$ and $P_I(T)=\sum_{m \in \mathcal{N}_f} \ket{m}\bra{m} + \sum_{o \in \mathcal{O}} \ket{o}\bra{o}$. So, we get
\begin{equation}
\begin{split}
Q &= \text{Tr}_{\mathcal{B}} (P_I(T)-P_I(0))\\
&= \text{Tr}_{\mathcal{B}}(\sum_{m \in \mathcal{N}_f} \ket{m}\bra{m} - \sum_{n \in \mathcal{N}_i} \ket{n}\bra{n})\\
&=\sum_{m \in \mathcal{N}_f} \text{Tr}_{\mathcal{B}} (\ket{m}\bra{m}) - \sum_{n \in \mathcal{N}_i} \text{Tr}_{\mathcal{B}} (\ket{n}\bra{n}).
\end{split}\label{quant_Q}
\end{equation}
If a state $\ket{n}$ is completely localized in $\mathcal{B}$, then $\text{Tr}_{\mathcal{B}} (\ket{n}\bra{n})=1$, and if it is completely localized in $\mathcal{S}-\mathcal{B}$, then $\text{Tr}_{\mathcal{B}} (\ket{n}\bra{n})=0$. So, if all states in $\mathcal{N}_f$ and $\mathcal{N}_i$ are completely localized either in $\mathcal{B}$ or in $\mathcal{S}-\mathcal{B}$, then $\sum_{m \in \mathcal{N}_f}\text{Tr}_{\mathcal{B}} (\ket{m}\bra{m})$ and $\sum_{n \in \mathcal{N}_i} \text{Tr}_{\mathcal{B}} (\ket{n}\bra{n})$ would be integers, giving rise to a quantized adiabatic charge $Q$. 
Now, if all states in $\mathcal{N}_i$ and $\mathcal{N}_f$ are completely localized in $\mathcal{S}-\mathcal{B}$, then $\sum_{m \in \mathcal{N}_f} \text{Tr}_{\mathcal{B}} (\ket{m}\bra{m})=\sum_{n \in \mathcal{N}_i} \text{Tr}_{\mathcal{B}} (\ket{n}\bra{n})=0$ and $Q=0$.  Also, if all states in $\mathcal{N}_i$ and $\mathcal{N}_f$ are completely localized in $\mathcal{B}$, then $\sum_{m \in \mathcal{N}_f} \text{Tr}_{\mathcal{B}} (\ket{m}\bra{m})=\sum_{n \in \mathcal{N}_i} \text{Tr}_{\mathcal{B}} (\ket{n}\bra{n})=N$ and $Q=0$. A non-trivial quantized contribution to the adiabatic charge transport is obtained when a pair of states, $\ket{m} \in \mathcal{N}_f$ and $\ket{n} \in \mathcal{N}_i$, are localized in such a way that one of them is completely localized in $\mathcal{B}$ and the other is completely localized in $\mathcal{S}-\mathcal{B}$.
\newline

\section{Condition for an edge-like state to be completely localized on the sites immediately enclosing a triangle of a given generation of SG-3 } \label{Complete_localization_condition}
We start with the Harper-Hofstadter Hamiltonian on SG-3, given by
\begin{equation}
\hat{H}=\sum_{\langle jk \rangle} H_{jk} c^{\dagger}_j c_k ,
\end{equation}
where $H_{jk}=-e^{-i\theta_{jk}}$, when the sites labeled by the indices $j$ and $k$ are nearest neighbors and $0$ otherwise. $\theta_{jk}$ is the same as defined in Eq.\ \eqref{HH} of the main text. Let ${H}$ denote the Hamiltonian matrix whose elements are $H_{ij}$. Consider a triangle of a given generation of SG-3. We put a flux tube, carrying flux $2\pi\varphi$, through this triangle. Let us denote the set of all sites immediately enclosing the triangle to be $\mathcal{P}$, and the set containing the rest of the sites to be $\mathcal{Q}$. Now consider an edge-like state $\ket{\psi}=\sum_{j} \psi_j \ket{\textbf{r}_j}$. By breaking into sectors of $\mathcal{P}$ and $\mathcal{Q}$, the Hamiltonian can be represented in the matrix  form as follows
\begin{equation}
H=
\begin{bmatrix}
H_{\mathcal{P}} & H_{\mathcal{PQ}}\\
H_{\mathcal{QP}} & H_{\mathcal{Q}}
\end{bmatrix},
\end{equation}
where $H_{\mathcal{P}_{jk}}=H_{jk}$, $\forall j,k \in \mathcal{P}$; $H_{\mathcal{PQ}_{jk}}=H_{jk},~ \forall j \in \mathcal{P},~ k \in \mathcal{Q}$; $H_{\mathcal{QP}_{jk}}=H_{jk}, ~\forall j \in \mathcal{Q},~ k \in \mathcal{P}$; and $H_{\mathcal{Q}_{jk}}=H_{jk}, ~\forall j,k \in \mathcal{Q}$. Similarly the state $\ket{\psi}$ can be expressed as

\begin{equation}
\ket{\psi}=\ket{\psi^{\mathcal{P}}}+\ket{\psi^{\mathcal{Q}}},
\end{equation}
where $\ket{\psi^{\mathcal{P}}}=\sum_{p \in \mathcal{P}} \psi_p \ket{\textbf{r}_p}$ and $\ket{\psi^{\mathcal{Q}}}=\sum_{q \in \mathcal{Q}} \psi_q \ket{\textbf{r}_q}$. In the vector form, let $\Psi^{\mathcal{P}}=[ \psi_{p_1} ~ \psi_{p_2}....\psi_{p_{N_p}} {]}^{\text{T}} ~ \forall \lbrace p_i \rbrace \in \mathcal{P}$ and $\Psi^{\mathcal{Q}}=[ \psi_{q_1} ~ \psi_{q_2}....\psi_{q_{N_q}} {]}^{\text{T}} ~ \forall \lbrace q_i \rbrace \in \mathcal{Q}$ be the representations for $\ket{\psi^{\mathcal{P}}}$ and $\ket{\psi^{\mathcal{Q}}}$ respectively. 
\newline

If the state is completely localized on sites in $\mathcal{P}$, then $\Psi^{\mathcal{Q}}=0$. Now, given that $\Psi=[ \Psi^{\mathcal{P}} ~ \Psi^{\mathcal{Q}} ]^{\text{T}}$ is an eigenstate of $H$, we get that 
\begin{equation}
H\Psi=\begin{bmatrix}
H_{\mathcal{P}} & H_{\mathcal{PQ}}\\
H_{\mathcal{QP}} & H_{\mathcal{Q}}
\end{bmatrix}
\begin{bmatrix}
\Psi^{\mathcal{P}} \\
0
\end{bmatrix}=
\begin{bmatrix}
H_{\mathcal{P}} \Psi^{\mathcal{P}}\\
H_{\mathcal{QP}} \Psi^{\mathcal{P}}
\end{bmatrix}=E_m
\begin{bmatrix}
\Psi^{\mathcal{P}}\\
0
\end{bmatrix}.
\end{equation}
This implies that $ \Psi^{\mathcal{P}}$ must be an eigenstate of $H_{\mathcal{P}}$ and $H_{\mathcal{QP}} \Psi^{\mathcal{P}}=0$. $\Psi_{\mathcal{P}}$ can be analytically determined. To do that, we first point out that the sites in $\mathcal{P}$, together with the bonds with their respective nearest neighbors in $\mathcal{P}$, form a ring. They can be indexed linearly from 1 to $N_{\mathcal{P}}$, where $N_{\mathcal{P}}$ is the total number of sites in $\mathcal{P}$. So $\Psi^{\mathcal{P}}$ can be written as $\Psi^{\mathcal{P}}=[ \psi_{1} ~ \psi_{2}...\psi_{p}...\psi_{{N_p}} {]}^{\text{T}}$.  We do a gauge transformation given by $c^{\dagger}_j \rightarrow {c'}^{\dagger}_j=e^{-i\Theta_j}c^{\dagger}_j$, where $\Theta_1=0$, $\Theta_j=\sum^{j}_{n=2} \theta_{n-1,n} - (j-1)2{\pi}\Phi/N_{\mathcal{P}}\phi_0$ for $j \in \{2,3,\ldots,N_{\mathcal{P}}\}$, where $2{\pi}\Phi$ is the total flux threaded through the area enclosed by the sites in $\mathcal{P}$. Under this transformation, $ \Psi^{\mathcal{P}} \rightarrow {\Psi'}^{\mathcal{P}}$ and $H_{\mathcal{P}} \rightarrow H'_{\mathcal{P}}$, where $H'_{\mathcal{P}}$ is a Hermitian circulant matrix given by
\begin{equation}
H'_{\mathcal{P}}= 
\begin{bmatrix}
0 & t & 0 & \cdots & 0 & t^*\\
t^* & 0 & t & \cdots & 0 & 0\\
0 & t^* & 0 & t & \cdots & 0 \\
\vdots & \vdots & \vdots & \ddots & \vdots & \vdots\\
t & 0 & 0 & \cdots & t^* & 0
\end{bmatrix},
\end{equation}
and $t=e^{-i{2\pi\Phi}/{N_{\mathcal{P}}{\phi_0}}}$. The eigenvectors of $H'_{\mathcal{P}}$ are given by $\psi({\kappa})=[~ \omega^{\kappa} ~ \omega^{2{\kappa}} ~ \omega^{3{\kappa}} \ldots \omega^{p{\kappa}} \ldots \omega^{N_{\mathcal{P}}{\kappa}} ]^{\text{T}} ~ \forall {\kappa} \in \{0,1,2,\ldots,N_{\mathcal{P}}-1\}$, where $\omega=e^{i2\pi / N_{\mathcal{P}}}$. So, ${\Psi'}^{\mathcal{P}}$ must be equal to $\psi({\kappa})$ for some ${\kappa} \in \{0,1,2,\ldots,N_{\mathcal{P}}-1\}$. Now, $\Psi^{\mathcal{P}}$ can be obtained by  inverting the gauge transform and so we get 
\begin{eqnarray}\label{analytical_psi}
\Psi^{\mathcal{P}}({\kappa})=[ \psi_{1}({\kappa}) ~ \psi_{2}({\kappa})....\psi_{{N_p}}({\kappa}) {]}^{\text{T}},\\
 \psi_{p}({\kappa})=e^{i\Theta_p}\omega^{p{\kappa}}=e^{i\Theta_p}e^{i2\pi p{\kappa}/ N_{\mathcal{P}}}.
\end{eqnarray}
\newline

Given the analytical form of $\Psi^{\mathcal{P}}$, we can plug Eq.\ \eqref{analytical_psi} into the equation, $H_{\mathcal{QP}} \Psi^{\mathcal{P}}=0$, and get 
\begin{equation}\label{neighbor condition}
\sum_{p \in \mathcal{P}}H_{\mathcal{QP}_{q,p}}\psi_{p}=0,~ \forall q \in \mathcal{Q}.
\end{equation}
Notice that every site $q \in \mathcal{Q}$ either has exactly two consecutive nearest neighbors in $\mathcal{P}$ or zero nearest neighbors in $\mathcal{P}$. For the sites in $\mathcal{Q}$ which have zero nearest neighbors in $\mathcal{P}$, $H_{\mathcal{QP}_{q,p}}=0~\forall p \in \mathcal{P}$ and Eq.\ \eqref{neighbor condition} is trivially satisfied. For the rest of the sites $q_0 \in \mathcal{Q}$, let us say sites $p_0 \in \mathcal{P}$ and $p_0+1 \in \mathcal{P}$ are its nearest neighbors. Then we have $H_{\mathcal{QP}_{q_0,p_0}} {\psi_{p_0}} + H_{\mathcal{QP}_{q_0,p_0+1}} {\psi_{p_0+1}}=0$, which implies
\begin{equation}
e^{-i\theta_{q_0,p_0}}e^{i\Theta_{p_0}}{\omega^{p_0 {\kappa}}}+e^{-i\theta_{q_0,p_0+1}}e^{i\Theta_{p_0+1}}{\omega^{(p_0+1){\kappa}}}=0.\label{precondition}
\end{equation}
Simplifying Eq.\ \eqref{precondition}, we get the following condition
\begin{equation}\label{precondition2}
1+{\omega}^{\kappa} e^{i(\Theta_{p_0+1}-\Theta_{p_0})}e^{-i(\theta_{q_0, p_0+1}-\theta_{q_0, p_0})}=0.
\end{equation}
\newline

From the choice of $\lbrace \Theta_j \rbrace$, we get $\Theta_{p_0+1}-\Theta_{p_0}=\theta_{p_0,p_0+1}-2\pi\Phi/N_{\mathcal{P}} \phi_0$. Also, $\theta_{q_0,p_0}+\theta_{p_0,p_0+1}-\theta_{q_0,p_0+1}=\theta_{q_0,p_0}+\theta_{p_0,p_0+1}+\theta_{p_0+1,q_0}=-2\pi\phi/ \phi_0$, which is nothing but the flux through the triangle whose vertices are sites $q_0, ~p_0$ and $p_0+1$. Plugging this in Eq.\ \eqref{precondition2}, we get
\begin{eqnarray}
1+\omega^{\kappa} e^{-i2\pi\Phi/\phi_0 N_{\mathcal{P}}}e^{-i2\pi\phi/ \phi_0}=0 \\
\implies \dfrac{i2\pi {\kappa}}{N_{\mathcal{P}}}-\dfrac{i2\pi\Phi}{ N_{\mathcal{P}}\phi_0}-\dfrac{i2\pi\phi} {\phi_0}=(2n+1)i\pi\\
\implies {\kappa} - \dfrac{\Phi}{\phi_0}- N_{\mathcal{P}}\dfrac{\phi}{\phi_0}=\dfrac{(2n+1)}{2}N_{\mathcal{P}},\label{condition1}
\end{eqnarray}
where $n \in \mathbb{Z}$. We can express the total flux though the area enclosed by the $\mathcal{P}$ sites as the sum of the flux due to the magnetic field and the flux through the flux tube, $2\pi\Phi/ \phi_0=2\pi\phi\Delta / \phi_0+2\pi\varphi / \phi_0$, where $\Delta$ is the ratio of the area of the region enclosed by the $\mathcal{P}$ sites and the area of the triangle whose vertices are the sites $q_0,~ p_0$ and $p_0+1$. Plugging this into Eq.\ \eqref{condition1}, we get
\begin{equation}\label{condition}
{\kappa}=\dfrac{(2n+1)}{2}N_{\mathcal{P}}+(\Delta+N_{\mathcal{P}})\dfrac{\phi}{\phi_0}+\dfrac{\varphi}{\phi_0}.
\end{equation}
\newline

For SG-3, we notice that $N_{\mathcal{P}}=3z$, where $z$ is the number of sites on one side of the triangle enclosed by the $\mathcal{P}$ sites, and $z$ is always an even number. $\Delta$ is a natural number as it can be expressed in terms of $z$ as, $\Delta=z^2+2z-2$. Therefore, we must have
\begin{eqnarray}
(\Delta + N_{\mathcal{P}})\dfrac{\phi}{\phi_0} + \dfrac{\varphi}{\phi_0}=Z\label{condition-varphi}\\
{\kappa}=\dfrac{(2n+1)}{2}N_{\mathcal{P}}+Z, \label{condition-k}
\end{eqnarray}
where $Z$ is an integer. From Eqs.\ \eqref{condition-varphi} and \eqref{condition-k}, we conclude that the state labeled by $\kappa$ is completely localized on the sites in $\mathcal{P}$ if we choose $\varphi$ such that Eq.\ \eqref{condition-varphi} is fulfilled for the value of $Z$ that produces the right $\kappa$ in Eq.\ \eqref{condition-k}. The resulting state is an eigenstate of the angular momentum operator with eigenvalue $\kappa$.
\newline

\bibliography{ref.bib}

\end{document}